\documentclass[a4paper,10pt]{article}
\usepackage{amssymb}
\usepackage[utf8]{inputenc}
\usepackage{amsmath}
\usepackage{amsfonts}
\usepackage{cite}
\usepackage{geometry}
\usepackage{color}
\usepackage{graphicx}
\usepackage{subcaption}

\begin{document}

\title{Infinite conformal symmetry and emergent chiral (super)fields \\
of topologically non-trivial configurations: \\
From Yang-Mills-Higgs to the Skyrme model}
\author{Fabrizio Canfora$^{1,2}$, Diego Hidalgo$^{3}$, Marcela Lagos$^{3}$,
Enzo Meneses$^{3}$, Aldo Vera$^{3}$ \\
$^{1}$ \textit{Universidad San Sebasti\'an, Facultad de Ingenier\'ia, Arquitectura y Diseño, sede Valdivia,}\\
\textit{General Lagos 1163, Valdivia 5110693, Chile} \\
$^{2}$\textit{Centro de Estudios Cient\'{\i}ficos (CECS), Casilla 1469,
Valdivia, Chile}\\
$^{3}$\textit{Instituto de Ciencias F\'{\i}sicas y Matematicas,}\\
\textit{\ Universidad Austral de Chile, Casilla 567, Valdivia, Chile}\\
{\small fabrizio.canfora@uss.cl, diego.hidalgo@uach.cl,
marcela.lagos@uach.cl, enzo.meneses@alumnos.uach.cl, aldo.vera@uach.cl}}
\maketitle

\begin{abstract}
The present manuscript discusses a remarkable phenomenon concerning
non-linear and non-integrable field theories in $(3+1)$-dimensions, living
at finite density and possessing non-trivial topological charges and
non-Abelian internal symmetries (both local and global). With suitable types
of ans$\overset{..}{a}$tze, one can construct infinite-dimensional families
of analytic solutions with non-vanishing topological charges (representing
the Baryonic number) labelled by both two integers numbers and by free
scalar fields in $(1+1)$-dimensions. These exact configurations represent $%
(3+1)$-dimensional topological solitons hosting $(1+1)$-dimensional chiral
modes localized at the energy density peaks. First, we analyze the
Yang-Mills-Higgs model, in which the fields depend on all the space-time
coordinates (to keep alive the topological Chern-Simons charge), but in such
a way to reduce the equations system to the field equations of
two-dimensional free massless chiral scalar fields. Then, we move to the
non-linear sigma model, showing that a suitable ansatz reduces the field
equations to the one of a two-dimensional free massless scalar field. Then,
we discuss the Skyrme model concluding that the inclusion of the Skyrme term
gives rise to a chiral two-dimensional free massless scalar field (instead
of a free massless field in two dimensions as in the non-linear sigma model)
describing analytically spatially modulated Hadronic layers and tubes. The
comparison of the present approach both with the instantons-dyons liquid
approach and with Lattice QCD is shortly outlined.
\end{abstract}

\newpage

\tableofcontents


\section{Introduction}


It is well known that in quantum chromodynamics (QCD) color confinement is
closely related to the existence of topologically non-trivial configurations
(see \cite{greensite,DualSC,manton,BaMa,shifman1,shifman2,
weinberg,NP,NP1,NP2,NP3,NP3.1,NP4,NP5} and references therein), while in the
ultraviolet sector quarks and gluons should be liberated \cite%
{cabibbo,AF1,AF2}. The great advances in lattice quantum chromodynamics
(LQCD henceforth) \cite{[15],[16],[19],[20],[21],[22],[23],[24]} can only
partially compensate the poor analytic control on such non-perturbative
issues arising in the phase diagram of non-Abelian gauge theories.
Therefore, many open problems would greatly benefit from the presence of
explicit solutions relevant to the phase diagram of QCD. In this paper, we
will present a concise list of tools, although these are also useful when
analyzing different kinds of questions.

An area in which the results and tools of LQCD badly need some further
analytic insights is the analysis of the phase diagram of QCD at finite (and
low) temperature and with Baryon chemical potential; one of the main
problematic issues being the infamous sign problem (see \cite{kogutBook} for
a detailed review). In this case, the methods of AdS/CFT are not especially
effective since, only at high enough temperatures, supersymmetric Yang-Mills
theory gets very close to Yang-Mills theory (see \cite{Yaffe, Yaffe2} and
references therein). Moreover, besides the huge theoretical interest in
achieving a deeper understanding of this region of the phase diagram, there
are many situations of high phenomenological interest (such as heavy-ion
collisions, quark-gluon plasma, neutron stars and so on) in which novel
analytic techniques would be extremely useful to complement the available
observations. Among these, one of the most relevant is the appearance of
regular-shaped structures at finite density (called nuclear pasta states;
see \cite%
{pasta1,pasta2,pasta2a,pasta2b,pasta3,pasta4,pasta5,pasta6,pasta7,pasta8,pasta9,pasta10}%
) and the important transport properties within them, whose numerical
treatment is quite challenging \cite%
{pastacond1,pastacond2,pastacond3,pastacond4,pastacond5}.

There are two obvious ways to analyze these issues. One can either begin
with the analysis of Yang-Mills theory (which is more fundamental), or one
can start directly with the non-linear sigma model (NLSM) and the Skyrme
model (which is the low energy limit of QCD at leading order in the large $%
N_{c}$ 't Hooft expansion \cite{skyrme,Witten,witten0,bala0,Bala1,ANW}).
These models, at first glance, are very different, as Yang-Mills theory is a
gauge theory while the NLSM and the Skyrme model only possess global
symmetries. Thus, one could think that these two possibilities should be
treated with different methods. Nevertheless, we will show that it is
possible to devise a unified strategy able to identify sectors of the $(3+1)$%
-dimensional theories that, at the same time, possess arbitrary Baryonic
charge as well as an infinite-dimensional conformal symmetry. It is quite
amusing that the only difference between the infinite-dimensional conformal
symmetry appearing in Yang-Mills theory and the NLSM on one side, and
Yang-Mills-Higgs and the Skyrme theory, on the other side, is that, in the
former cases, one gets an effective two-dimensional conformal field theory
(CFT) while in Yang-Mills-Higgs and the Skyrme cases, one gets a
two-dimensional chiral CFT. This intriguing result could be related to the
fact that the Skyrme theory is the low energy limit of QCD (and not just of
Yang-Mills theory) and knows about chiral symmetry breaking. Needless to
say, the possibility to use the tools of two-dimensional CFT in $(3+1)$%
-dimensional theories (such as Yang-Mills and Skyrme, which are the
prototypes of non-linear and non-integrable field theories) open unexpected
and novel perspectives on the analysis of the phase diagram at finite
temperature and chemical potential.

A systematic tool to construct non-spherical hedgehog ansatz suitable to
describe finite density effects that have been developed in Refs. \cite%
{56,56a, 56b,56c, 56d, 56e, gaugsk, LastUS1, LastUS2, LastUS3} for the
Skyrme model, and for the Einstein-Yang-Mills case in Refs. \cite%
{ourYM1,ourYM2,ourYM3}. In the present article, we will further generalize these
results to extend the space of analytical solutions (and the tools that allow
obtaining relevant physical information of these systems of topological
solitons defined in a $(3+1)$-dimensional finite volume) disclosing the appearance of chiral
conformal degrees of freedom representing modulations of Hadronic tubes and layers. Although, in the
present the paper, we will not discuss the coupling with gravity of the NLSM
and Yang-Mills theories, there are already quite a few examples in the
literature which shows that the current approach is convenient even when the
coupling with general relativity is taken into account (see, for instance, 
\cite{grav1,grav2,grav3,grav4,grav5,grav6,grav7}).

\subsection{About the new analytical solutions}


The considerable interest in constructing analytic solutions in theories
with non-Abelian internal symmetries (both local and global) and non-trivial
topological charges arises from the fact that in all the theories admitting
topological solitons, such charges have a profound physical meaning (such as
the Baryonic charge, as will be discussed in the following sections). As far
as the phase diagram is concerned, it is crucial to analyze what happens
when a finite amount of topological charge is forced to live within a
limited spatial volume. In this case, practical analytic tools are extremely
welcome due, for instance, to the sign problem. On the other hand, the
common belief is that it is impossible to develop such tools for (at least)
two reasons. Firstly, one necessarily has to abandon spherically symmetric
ans$\overset{..}{a}$tze for the fields. Secondly and quite generically, the
requirement of a non-vanishing topological charge increases the complexity
of the field equations to be solved since a non-vanishing topological
density implies that there must be at least three independent degrees of
freedom depending non-trivially on three different spatial coordinates in $%
(3+1)$-dimensions. Resuming:

\begin{enumerate}
\item the \textit{departure from spherical symmetry} (generated by the
presence of \textquotedblleft a box" within which the solitons are forced to
live), together with

\item the \textit{requirement of a non-vanishing topological charge},
\end{enumerate}

reduce considerably the possibility to derive analytic results on the phase
diagram of topologically non-trivial configurations of theories such as
Yang-Mills, NLSM and the Skyrme model. One could reason as follows: the
analytic tools of two-dimensional CFT would be handy and welcome in
analyzing the phase diagram of $(3+1)$-dimensional Yang-Mills-Higgs theory
(or Skyrme model) due to the difficulties analyzing it even with LQCD. Then, 
\textit{why do not we assume that the main fields} (either $A_{\mu }$ for
Yang-Mills or $U\in SU(2)$ for the NLSM and Skyrme theory) \textit{only
depend on one spatial coordinate and on time (so that one could hope to use
some \textquotedblleft two-dimensional CFT technologies")?}

The answer is that such a naive approach would fail. \textit{First of all},
the topological charge (to be defined in the following sections) would
vanish identically, so that one would gain no information about the phase
diagram at finite Baryon density. \textit{Moreover}, already the head-on
collision of (topologically trivial) plane waves depending on only two
coordinates is intractable from the analytic viewpoint, and numerical
methods must be used\footnote{%
For instance, already analysis of head-on collisions of $(1+1)$-dimensional
kinks, which is far simpler than Yang-Mills theory in $(3+1)$ dimensions,
can only be dealt numerically \cite{Kink1, Kink2}.} (see \cite%
{collision,collision00, collision0, collision1, collision2, collision3,
collision5} and references therein). Hence, at first glance, one might argue
that the analytic study of dynamical processes involving solitonic
configurations with non-vanishing topological charge in $(3+1)$-dimensions
is not feasible.

In fact, here we will show that, from the analytic viewpoint, the above two
circumstances (namely, the need to depart from spherical symmetry and the
necessity to keep alive the topological charge) are an opportunity rather
than an obstruction. The tools to be developed here give rise, among other
things, to genuine $(3+1)$-dimensional non-homogeneous exact
solutions representing spatially modulated Hadronic layers and tubes,
allowing to estimate their contributions to the partition function at low
temperatures and Baryon chemical potential and also to compute relevant
quantities.

\subsection{Notation and conventions}


In this work, we will use the following convention. Greek indices run over
the space-time dimensions with mostly plus signature, and Latin indices are
reserved for those of the internal space. Also, we work in natural units,
such that the Boltzmann's constant $k_{\text{B}}$, the reduced Planck's
constant $\hbar$, and the speed of light $c$ are set to one.

As we are interested in studying topological solitons at finite volume, we
will use the metric of a box, which in $(3+1)$-space-time dimensions reads 
\begin{equation}
ds^{2}=-dt^{2}+L_{r}^{2}\,dr^{2}+L_{\theta }^{2}\,d\theta ^{2}+L_{\phi
}^{2}\,d\phi ^{2}\ ,  \label{Box4}
\end{equation}%
where $\{r,\theta ,\phi \}$ are Cartesian dimensionless coordinates whose
ranges will be defined in each case, $\{L_{r},L_{\theta },L_{\phi }\}$ are
constants with dimension of length that fix the volume of the box in which
the solitons are confined and $g=-L_{r}^{2}L_{\theta }^{2}L_{\phi }^{2}$
will denote the metric determinant. Also we denote $\nabla _{\mu }$ as the
Levi-Civita covariant derivative constructed with the Christoffel symbols, $%
\partial _{\mu }$ as the partial derivative, and the covariant derivative, $%
D_{\mu }$, acts as 
\begin{equation}
D_{\mu }(\cdot )=\partial _{\mu }(\cdot )+[A_{\mu },\,\cdot \,]\ ,  \label{D}
\end{equation}%
with $A_{\mu }$ the components of the non-Abelian connection. We will
consider as internal symmetry group the $SU(2)$ Lie group\footnote{%
Here we will consider the $SU(2)$ case but the present results can be
extended to the $SU(N)$ case.}, whose generators are 
\begin{equation}
\mathbf{t}_{k}=i\,\sigma _{k}\ ,  \label{ti}
\end{equation}%
being $\sigma _{k}$ the Pauli matrices. The matrices $\mathbf{t}_{i}$
satisfy the relation 
\begin{equation}
\mathbf{t}_{i}\mathbf{t}_{j}=-\delta _{ij}\mathbf{1}_{2}-\epsilon _{ijk}%
\mathbf{t}_{k}\ ,
\end{equation}%
where $\mathbf{1}_{2}$ is the $2\times 2$ identity matrix, $\delta _{ij}$
the Kronecker delta and $\epsilon _{ijk}$ the totally antisymmetric
Levi-Civita symbol.

The fundamental field of the Yang-Mills theory, namely the non-Abelian
connection $A$, splits as 
\begin{equation}  \label{A}
A=A_{\mu }^{j}\mathbf{t}_jdx^{\mu } \ ,
\end{equation}
while the fundamental field of the NLSM and the Skyrme model is the scalar
field $U(x)\in SU(2)$, so that 
\begin{equation}  \label{R}
R_{\mu} = U^{-1} \partial_{\mu} U = R_{\mu}^{j} \mathbf{t}_j \ ,
\end{equation}
is in the $\mathfrak{su}(2)$ algebra.

The relevant topological properties of the solutions that we will construct
in this work are encoded in the Chern-Simons (CS) density (for the
Yang-Mills theory) and in the Baryon charge density (for the NLSMs). These
are given, respectively, by 
\begin{equation}
\rho _{\text{CS}}=J_{0}^{\text{CS}} \ , \qquad \text{where} \qquad J_{\mu }^{%
\text{CS}}=\frac{1}{8\pi ^{2}}\varepsilon _{\mu \nu \rho \sigma }\text{Tr}%
\left( A^{\nu }\partial ^{\rho }A^{\sigma }+\frac{2}{3}A^{\nu }A^{\rho
}A^{\sigma }\right) \ ,  \label{CScurrent}
\end{equation}%
\begin{equation}
\rho _{\text{B}} = \frac{1}{24\pi^2}\left( U^{-1}\partial U\right)
^{3}\equiv \frac{1}{24\pi^2}\varepsilon _{ijk} \text{Tr}\left\{ \left(
U^{-1}\partial ^{i}U\right) \left( U^{-1}\partial ^{j}U\right) \left(
U^{-1}\partial ^{k}U\right) \right\} \ .  \label{rhoB}
\end{equation}%
The integral of the above densities over a space-like hypersurface
represents the CS charge and the Baryonic charge of the corresponding
configurations, 
\begin{equation}  \label{charges}
Q_{\text{CS}}=\int \rho_{\text{CS}}~ dV \,, ~~~ \qquad B=\int \rho_{\text{B}}~
dV \ .
\end{equation}
The paper is organized as follows: In Section 2, we study the Yang-Mills theory in $(3+1)$-dimensions showing that, with an appropriate ansatz, the field equations are reduced to that of a two-dimensional free massless scalar field. We also offer that the inclusion of a Higgs field converts the resulting CFT into a chiral theory. In Section 3, we move to the study of NLSM in $(3+1)$-dimensions, showing that the theory can be reduced to a two-dimensional CFT. In Section 4, we show that the inclusion of the Skyrme term in the NLSM defines a chiral CFT for two types of configurations describing nuclear pasta states. In Section 5, we study the phase diagram and the contribution of the partition functions of the analytic topological solitons. The final section is dedicated to conclusions.


\section{Yang-Mills-Higgs theory in $(3+1)$-dimensions}


In this section, before moving to the Yang-Mills-Higgs case (which has not
been analyzed previously in the literature), we will study the Yang-Mills
theory in $(3+1)$-dimensions reviewing the results in Ref. \cite{ourYM4l},
showing how the field equations can be reduced to that of a two-dimensional
free massless scalar field in $(1+1)$-dimensions keeping alive the
topological charge. The concepts introduced here will be helpful also in the
following sections, where we will show that a similar construction can also
be carried out on NLSMs.

\subsection{Conformal field theory in two dimensions from pure Yang-Mills theory}


The Yang-Mills theory in $(3+1)$-dimensions is described by the action 
\begin{equation}
I[A]=\frac{1}{2e^{2}}\int d^{4}x\sqrt{-g}\, \text{Tr}(F_{\mu \nu }F^{\mu \nu
})\ ,  \label{I}
\end{equation}%
where $e$ is the Yang-Mills coupling constant, and the field strength
components $F_{\mu \nu}$ are defined in terms of the non-Abelian connection $%
A_\mu$ as 
\begin{equation*}
F_{\mu \nu }=\partial _{\mu }A_{\nu }-\partial _{\nu }A_{\mu }+[A_{\mu
},A_{\nu }] \,.
\end{equation*}%
The field equations of the theory, obtained varying the action in Eq.~%
\eqref{I} with respect to the fundamental field $A_{\mu}$, are 
\begin{equation}
\nabla _{\nu }F^{\mu \nu }+[A_{\nu },F^{\mu \nu }]=\ 0\ ,  \label{yme0}
\end{equation}%
while the energy-momentum tensor of the theory turns out to be 
\begin{equation}  \label{TmunuYM}
T_{\mu \nu }=-\frac{2}{e^{2}}\text{Tr}\biggl(F_{\mu \alpha }F_{\nu
}{}^{\alpha }-\frac{1}{4}g_{\mu \nu }F_{\alpha \beta }F^{\alpha \beta }%
\biggl)\ .
\end{equation}
One of the main goals of this paper is to construct a formalism able to
describe how topologically non-trivial configurations react when they are
forced to live within a finite box; this issue must be addressed in the
finite density analysis.

The easiest way to take into account finite volume effects is to use the
flat metric defined in Eq.~\eqref{Box4}, with the ranges 
\begin{equation}
0\leq \theta \leq 2\pi \ ,\qquad 0\leq \phi \leq \pi \ ,\qquad 0\leq r\leq
4\pi \ .\   \label{ranges1}
\end{equation}%
The above ranges for the coordinates $\theta $, $\phi $ and $r$ are related
to the Euler angle parameterization for $SU(2)$ valued fields. Let us define
the following $U(x)\in SU(2)$ field 
\begin{equation}
U=\exp \left( p\,\theta \frac{\mathbf{t}_{3}}{2}\right) \exp \left( H\left(
t,\phi \right) \frac{\mathbf{t}_{2}}{2}\right) \exp \left( q\,r\frac{\mathbf{%
t}_{3}}{2}\right) \,,  \label{ans2}
\end{equation}%
where $p$ and $q$ are non-vanishing integers\footnote{%
There will be one more restriction on $p$ and $q$ that will be discussed
later on.}. The theory of Euler angles for $SU(N)$ \cite%
{euler1,euler2,euler3}, tells that, when $p$ and $q$ are non-vanishing
integers, the range of $\theta $ (appearing in the left factor of the
decomposition in Eq.~\eqref{ans2}) and the range of $r$ (appearing in the
right factor of the decomposition in Eq.~\eqref{ans2}) must be as in Eq.~%
\eqref{ranges1}. As far as the central factor $H\left( t,\phi \right) $ is
concerned, there are two options. If the field $H\left( t,\phi \right) $
satisfies periodic boundary conditions 
\begin{equation}
H\left( t,\phi =0\right) =H_{0}=H\left( t,\phi =\pi \right) \,,
\label{ans2.01}
\end{equation}%
the CS charge vanishes\footnote{%
Although the CS density can still be non-trivial.}. The other boundary
condition for $H\left( t,\phi \right) $ arises naturally taking into account
that $H\left( t,\phi \right) $ appears in the central factor of the Euler
angles decomposition of an $SU(2)$ element (see \cite{euler1,euler2, euler3}%
), 
\begin{equation}
H\left( t,\phi =0\right) =0\,,~~~~H\left( t,\phi =\pi \right) =\pi \ ,
\label{ans2.1}
\end{equation}%
or%
\begin{equation*}
H\left( t,\phi =0\right) =\pi \ ,\ \ H\left( t,\phi =\pi \right) =0\ .
\end{equation*}%
The option here above ensures that the $SU(2)$-valued element $U$ defined in
Eqs.~\eqref{ranges1}, (\ref{ans2}) and (\ref{ans2.1}) wraps an integer
number of times around the group manifold of $SU(2)$; in other words, $U$
has a non-vanishing winding number. In this case, both the CS charge and the
CS density in Eq.~\eqref{CScurrent} associated with the gauge field will be
non-trivial. It is well known that $\rho _{\text{CS}}$ defined in Eq.~%
\eqref{CScurrent}) is the ``non-perturbatively induced Baryonic charge" of
the gauge configuration \cite{anomalies3} (see also \cite{anomalies,
anomalies1, anomalies2} and references therein).

In order to find an ansatz such that $\rho_{\text{CS}}$ defined in Eq.~%
\eqref{CScurrent} will be non-zero and, at the same time, the field
equations can be solved analytically, one can follow Refs. \cite{56a, 56b}, 
\cite{LastUS2, LastUS3, ourYM1, ourYM2, ourYM3}, \cite{ourYM4l} and \cite%
{Daniel}, arriving to the following form for the Yang-Mills potential 
\begin{equation}
A_{\mu }=\sum_{j=1}^{3}\lambda _{j}\Omega _{\mu }^{j}\mathbf{t}_{j}\,,\qquad
U^{-1}\partial _{\mu }U=\sum_{j=1}^{3}\Omega _{\mu }^{j}\mathbf{t}_{j}\ ,
\label{ans1}
\end{equation}%
where $H(t,\phi )$ in Eq. \eqref{ans2} and the $\lambda _{i}$ functions in
Eq.~\eqref{ans1} are explicitly given by 
\begin{equation}
H\left( t,\phi \right) =\arccos \left( G\right) \,,~~G=G\left( t,\phi
\right) \ ,  \label{ans3}
\end{equation}%
\begin{eqnarray}
\lambda _{1}\left( t,\phi \right) &=&\lambda _{2}\left( t,\phi \right) =%
\frac{G}{\sqrt{G^{2}+\exp (2\eta )}}\overset{def}{=}\lambda \left( t,\phi
\right) \ ,\ \ \ \lambda _{3}\left( t,\phi \right) =1\ ,\ \eta \in 
\mathbb{R}
\ ,  \label{ans4} \\
G\left( t,\phi \right) &=&\exp (3\eta )\frac{F}{\sqrt{1-\exp (4\eta )\cdot
F^{2}}}\,,~~~~F=F\left( t,\phi \right) \ .  \label{ans4.1}
\end{eqnarray}%
The real parameter $\eta $ will be fixed by requiring that the CS charge is
an integer.

The option in Eq.~\eqref{ans2.01} gives rise to the following boundary
condition for $F\left( t,\phi \right) $, 
\begin{equation}
F\left( t,\phi =0\right) =F_{0}=F\left( t,\phi =\pi \right) \ .
\label{ans2.002}
\end{equation}%
In the latter case the CS charge vanishes. On the other hand, the option in
Eq.~\eqref{ans2.1}, in terms of $F\left( t,\phi \right) $,\ reads 
\begin{equation}
F\left( t,\phi =0\right) =-\frac{\exp (-2\eta )}{\sqrt{1+\exp (2\eta )}}\
,\qquad F\left( t,\phi =\pi \right) =\frac{\exp (-2\eta )}{\sqrt{1+\exp
(2\eta )}}\ ,  \label{BC}
\end{equation}%
in order to have a non-zero CS charge. In this case both the CS charge and
the CS density will be non-trivial. Then, we say that this configuration is
topologically non-trivial.

The components of the gauge field can be easily computed taking into account
the well known expression of the $\Omega _{\mu }^{j}$\ in the case of the
Euler parameterization. Thus, explicitly, $A_{\mu }$ reads%
\begin{align}
A_{\mu }=&\lambda \left( t,\phi \right) \biggl[ \frac{\mathbf{t}_{1}}{2}
\left\{ -\sin \left( qr\right) dH+p\cos \left( qr\right) \sin \left(
H\right) d\theta \right\}   + \frac{\mathbf{t}_{2}}{2}\{ \cos \left( qr\right) dH+p\sin \left(
qr\right) \sin \left( H\right) d\theta \} \biggl] \notag\\
&+ ~\frac{\mathbf{t}_{3}}{2}%
\left[ qdr+p\cos (H)d\theta \right] \,,  \label{ansgauge}
\end{align}
where 
\begin{equation*}
dH=\frac{\partial H}{\partial t}dt+\frac{\partial H}{\partial \phi }d\phi \ .
\end{equation*}
The fact that $d\lambda \wedge dH=0$, together with the gradients of the coordinates $r$, $\theta $
and $\phi $\ are mutually orthogonal, simplifies many of the computations. The above ansatz is the key
to getting the paper's main results, and the rest is a direct computation.

With the above, the complete set of $(3+1)$-dimensional Yang-Mills field
equations with the ansatz in Eqs.~\eqref{ans2},~\eqref{ans1}, \eqref{ans3}, %
\eqref{ans4} and \eqref{ans4.1} reduces to 
\begin{equation}
\square F\, \equiv \, \left( \frac{\partial ^{2}}{\partial t^{2}}-\frac{1}{%
L_{\phi}^{2}}\frac{\partial ^{2}}{\partial \phi ^{2}}\right) F\, = \, 0\,,
\label{CFT1}
\end{equation}
which corresponds to the field equation of a free massless scalar field in
two dimensions.

\subsection{Energy-momentum tensor and topological charge}


A direct computation reveals that the topological density for the solution
defined above is given by 
\begin{equation}
\rho _{\text{CS}}=\frac{pq\exp (3\eta )}{16\pi ^{2}\left( 1-\exp (4\eta
)F^{2}\right) ^{3/2}}\frac{\partial F}{\partial \phi }\,,  \label{topdensity}
\end{equation}%
which is non-vanishing, as long as $\frac{\partial F}{\partial \phi }\neq 0$%
. On the other hand, the energy density, $T_{tt}$, and the on-shell
Lagrangian, $L_{\text{on-shell}}$, read, respectively, 
\begin{align}
T_{tt} =&\frac{p^{2}}{e^{2}L_{\theta}^{2}}\exp (5\eta )\cosh \left( \eta
\right) \left[ \left( \frac{\partial F}{\partial t}\right) ^{2}+\frac{1}{%
L_{\phi}^{2}}\left( \frac{\partial F}{\partial \phi }\right) ^{2}\right] \ ,
\label{T00} \\
L_{\text{on-shell}} =&\frac{p^{2}}{e^{2} L_{\theta}^{2}}\exp (5\eta )\cosh
\left( \eta \right) \left[ \left( \frac{\partial F}{\partial t}\right) ^{2}-%
\frac{1}{L_{\phi}^{2}}\left( \frac{\partial F}{\partial \phi }\right) ^{2}%
\right] \ .  \label{lonshell}
\end{align}
The full energy-momentum tensor, reads 
\begin{equation*}
T_{\mu \nu }=\left[ 
\begin{array}{cccc}
T_{tt} & 0 & 0 & P_{\phi } \\ 
0 & T_{rr} & 0 & 0 \\ 
0 & 0 & T_{\theta \theta } & 0 \\ 
P_{\phi } & 0 & 0 & T_{\phi \phi }%
\end{array}%
\right] \,,
\end{equation*}%
where%
\begin{align}
T_{rr} =& \frac{p^{2}L_r^2}{e^{2}L_{\theta}^{2}}\exp \left( 5\eta \right)
\cosh \left( \eta \right) \left[ \left( \frac{\partial F}{\partial t}\right)
^{2}-\frac{1}{L_\phi^{2}}\left( \frac{\partial F}{\partial \phi }\right) ^{2}%
\right] =- \frac{L_r^2}{L_{\theta}^{2}} T_{\theta \theta }\ ,  \label{tmunu1}
\\
T_{\phi \phi } =& \frac{p^{2}L_{\phi}^{2}}{e^{2}L_{\theta}^{2}}\exp \left(
5\eta \right) \cosh \left( \eta \right) \left[ \left( \frac{\partial F}{%
\partial t}\right) ^{2}+\frac{1}{L_{\phi}^{2}}\left( \frac{\partial F}{%
\partial \phi }\right) ^{2}\right] \,,  \label{tmunu2}
\end{align}%
\begin{equation}
T_{t\phi }=P_{\phi}=\frac{2 p^{2}\exp \left( 5\eta \right) \cosh \left( \eta
\right)}{e^{2} L_{\theta}^{2} }\frac{\partial F}{\partial t}\frac{\partial F%
}{\partial \phi }\,.  \label{tmunu3}
\end{equation}%
From the above, one can easily verify that the energy-momentum tensor is
traceless; $g^{\mu \nu }T_{\mu \nu }=0$, as it should be in Yang-Mills
theory in $(3+1)$-dimensions. It is also interesting to note that if one
``eliminates" the coordinates $r$ and $\theta $, the resulting
two-dimensional energy-momentum tensor in the $t$ and $\phi $\ directions is
still traceless (as it happens for a two-dimensional CFT). Explicitly, one
can take $T_{ab}$ defined as 
\begin{equation*}
T_{ab}=\left( 
\begin{array}{cc}
T_{tt} & P_{\phi } \\ 
P_{\phi } & T_{\phi \phi }%
\end{array}%
\right) \ ,\ \qquad a,b=t,\phi \,,
\end{equation*}%
as the effective energy-momentum tensor associated to the massless
two-dimensional scalar field $F$. As it is clear from Eq.~\eqref{topdensity}%
, the CS density associated to $F_{+}+F_{-}$ (where $F_{+}$ and $F_{-}$ are
the left and right movers mode expansion defined explicitly below) is the
sum of the topological charge density associated to $F_{+}$ plus to one
associated to $F_{-}$ only for small amplitudes, namely when 
\begin{equation}
\left\vert \exp \left( 4\eta \right) F\left( t,\phi \right) ^{2}\right\vert
\ll 1\,.  \label{cut-off1}
\end{equation}
On the other hand, when the temperature is high enough, it is natural to expect that the thermal fluctuations of $F\left( t,\phi \right) $ violate the above condition. That is why the CS density of these configurations (which can be interpreted as Baryonic charge density) is only well defined below a specific temperature.

The CS charge reads 
\begin{equation}
Q_{\text{CS}}=\frac{pq\exp (3\eta )}{2}\left. \left[ \frac{F}{\sqrt{1-\exp
(4\eta )F^{2}}}\right] \right\vert _{F(t,0)}^{F(t,\pi )}\,.
\label{gluonicBPS4}
\end{equation}%
As it has been already discussed, when $F(t,0)=F(t,\pi )$ the topological
charge vanishes. Thus, let us consider the boundary conditions for $F(t,\phi
)$ in Eq.~\eqref{BC}. The requirement to have an integer topological charge
can be expressed as follows. Introducing the useful auxiliary function 
\begin{equation}
\Omega (\eta ,a,b)\equiv \frac{\exp (3\eta )}{2}\left[ \frac{a}{\sqrt{1-\exp
(4\eta )a^{2}}}-\frac{b}{\sqrt{1-\exp (4\eta )b^{2}}}\right] \ ,
\label{aux1}
\end{equation}%
the topological charge reads%
\begin{equation*}
Q_{\text{CS}}=pq\cdot \Omega \left( \eta ,a=F(t,\pi ),b=F(t,0)\right) \ .
\end{equation*}%
Taking into account the boundary conditions for $F(t,\phi )$ in Eq.~%
\eqref{BC}, the quantity $\Omega \left( \eta ,a=F(t,\pi ),b=F(t,0)\right) $
can be further simplified, so that one arrives at the following expression
for the topological charge 
\begin{equation}
Q_{\text{CS}}\ =\ pq\ .  \label{aux2}
\end{equation}%
Consequently, in order to have integer topological charge, the number $pq$
must be integer. Here it is worth emphasizing that, although the field
equations in terms of $F\left( t,\phi \right) $\ are linear, an important
non-linear effect is manifest in Eqs. (\ref{topdensity}), (\ref{gluonicBPS4}%
) and (\ref{cut-off1}). Indeed, in order for the CS density in Eq.~(\ref%
{topdensity}) to be everywhere well defined, one must require 
\begin{equation}
\left\vert \exp \left( 4\eta \right) F\left( t,\phi \right) ^{2}\right\vert
\leq 1\,.  \label{cutoff1.1}
\end{equation}%
Since the thermal expectation value of $F\left( t,\phi \right) ^{2}$ grows
with temperature, the condition here above implies that the partition
function associated to the present family of exact solutions will be
well-defined only below a certain critical temperature beyond which the CS
density is not well defined anymore.

\subsection{Semi-classical considerations}


Let us remind the usual mode expansion of the solutions of Eq.~\eqref{CFT1}.
These can be written as 
\begin{equation}
F_{+}=\phi _{0}^{+}+v_{+}\left( \frac{t}{L_{\phi}}+\phi \right) +\sum_{n\neq
0}\left( a_{n}^{+}\sin \left[ n\left( \frac{t}{L_{\phi}}+\phi \right) \right]
+b_{n}^{+}\cos \left[ n\left( \frac{t}{L_{\phi}}+\phi \right) \right]
\right) \ ,  \label{general1}
\end{equation}%
\begin{equation}
F_{-}=\phi _{0}^{-}+v_{-}\left( \frac{t}{L_{\phi}}-\phi \right) +\sum_{n\neq
0}\left( a_{n}^{-}\sin \left[ n\left( \frac{t}{L_{\phi}}-\phi \right) \right]
+b_{n}^{-}\cos \left[ n\left( \frac{t}{L_{\phi}}-\phi \right) \right]
\right) \ ,  \label{general2}
\end{equation}%
where, as usual, $F_{+}$ refers to the left movers and $F_{-}$ to the right
movers ($v_{\pm }$ and $\phi _{0}^{\pm }$\ being integration constants,
which must satisfy three constraints that will be discussed below). Hence,
the most general topologically non-trivial configuration of the present
sector arises replacing $F=F_{+}+F_{-}$ into Eqs. (\ref{ans2}), (\ref{ans1}%
), (\ref{ans3}), (\ref{ans4}) and (\ref{ans4.1}). In order to have a clear
physical picture of the composition of solutions, it is convenient to choose 
$a_{n}^{\pm }$ and $b_{n}^{\pm }$ in such a way that 
\begin{equation*}
\widetilde{F}(t,\phi =0)=\widetilde{F}(t,\phi =\pi )=0\ ,
\end{equation*}%
where $\widetilde{F}(t,\phi )$ is the part of $F=F_{+}+F_{-}$ coming from
the sum over the integers $n$ in Eqs. (\ref{general1}) and (\ref{general2}).
Therefore, the topological charge in Eq. (\ref{gluonicBPS4}) is non-zero when%
\begin{equation*}
v_{+}-v_{-}\neq 0\ .
\end{equation*}%
In particular, $v_{\pm }$ and $\phi _{0}^{\pm }$\ in Eqs. (\ref{general1})
and (\ref{general2}) must be chosen as%
\begin{equation*}
F\left( t,\phi =0 \right) =\phi _{0}^{+}+\phi _{0}^{-}+\left(
v_{+}+v_{-}\right) \frac{t}{L_{\phi}}=\frac{\exp \left( -2\eta \right) }{%
\sqrt{1+\exp \left( 2\eta \right) }}
\end{equation*}%
\begin{equation}
\Rightarrow ~~ v_{+}+v_{-}=0\ ,\quad \phi _{0}^{+}+\phi _{0}^{-}=\frac{\exp
\left( -2\eta \right) }{\sqrt{1+\exp \left( 2\eta \right) }}\ \ ,
\label{bcond1}
\end{equation}%
\begin{equation*}
F\left( t,\phi =\pi \right) =\frac{\exp \left( -2\eta \right) }{\sqrt{1+\exp
\left( 2\eta \right) }}+\left( v_{+}-v_{-}\right) \pi =-\frac{\exp \left(
-2\eta \right) }{\sqrt{1+\exp \left( 2\eta \right) }}
\end{equation*}%
\begin{equation}
\Rightarrow ~~ v_{-}=\frac{\exp \left( -2\eta \right) }{\pi \sqrt{1+\exp
\left( 2\eta \right) }}\ .  \label{bcond2}
\end{equation}%
At a classical level, this is the most straightforward choice of boundary
conditions since it identifies which terms are responsible for the
topological charge and which are not.

At the semi-classical level, it is very tempting to introduce creation and
annihilation operators quantization corresponding to the above
mode-expansion, as it is usually done in quantising a free two-dimensional
scalar field. However, there are some intriguing differences.

First, in Eqs.~\eqref{general1} and \eqref{general2}, any term in
the expansion corresponds to an exact solution of the $(3+1)$-dimensional
Yang-Mills equations and not just to a solution of the linearized field
equations. Therefore, the Bosonic quantum operators $\alpha _{n}^{+}$, $%
\left( \alpha _{m}^{+}\right) ^{\dag }$ and $\alpha _{n}^{-}$, $\left(
\alpha _{m}^{-}\right) ^{\dag }$ (which are annihilation and creation
operators for the left and right movers, satisfying the obvious commutation
relations; see \cite{GSW}) create exact solutions of the semiclassical
Yang-Mills equations. This situation should be compared with the usual case
in which, given a particular solution of the $(3+1)$-dimensional Yang-Mills
equations, the small fluctuations (both at classical and quantum level)
around the given classical configurations are solutions of the linearized
field equations (while are not solutions of the exact field equations,
unless, of course, the theory is just a free theory).

Second, the constant terms $\phi _{0}^{\pm }$\ as well as the
linear terms in $t$ and $\phi $\ play an important role. According to Refs. 
\cite{anomalies,anomalies1, anomalies2,anomalies3}, the topological charge
can be interpreted as the Baryonic charge of the configuration. If this
interpretation is accepted, when the topological charge is odd, the
configuration is a Fermion, while when it is even, the configuration is a
Boson. This observation has no consequences for the operators $\left(\alpha
_{n}^{\pm }\ ,\ \left( \alpha_{n^{\prime }}^{\pm }\right) ^{\dag }\right) $
since these operators are Bosonic (due to the corresponding classical
solutions do not contribute to the topological charge). On the other hand,
the creation and annihilation operators associated with the solution's
linear part create a Boson or a Fermion depending on whether the topological
charge is even or odd. Hence, it is tempting to quantize $\phi _{0}^{\pm }$
and $v_{\pm }$ with commutators or anticommutators depending on the value of
the topological charge.

\subsection{Chiral conformal field theory from Yang-Mills-Higgs theory}


Now we will show that the construction presented above can be directly
generalized to the Yang-Mills-Higgs theory, but with the notable difference
that, this time, the theory is reduced to a chiral CFT in $(1+1)$-dimensions
instead of just a CFT.

The Yang-Mills-Higgs theory in $(3+1)$-dimensions is defined by the action 
\begin{equation}
I[A,\varphi]=\int d^{4}x\sqrt{-g}\,\biggl(\frac{1}{2e^{2}}\text{Tr}(F_{\mu
\nu }F^{\mu \nu})+\frac{1}{4} \text{Tr}(D_{\mu }\varphi D^{\mu }\varphi) %
\biggl) \,.  \label{IH2a}
\end{equation}
Here $\varphi$ is the Higgs field in the adjoint representation, and the
covariant derivative $D_{\mu}$ has been defined in Eq.~\eqref{D}. Varying
the action with respect to the fields $A_\mu$ and $\varphi$ we obtain the
field equations of the Yang-Mills-Higgs theory 
\begin{gather}  \label{Eq2a}
\nabla _{\nu }F^{\mu \nu }+[A_{\nu },F^{\mu \nu }]+\frac{e^{2} }{4}[\varphi
,D^{\mu }\varphi ]\ =\ 0\ , \\
D_{\mu }D^{\mu }\varphi \ =\ 0\ .  \label{EqHa}
\end{gather}
On the other hand, the energy-momentum tensor is 
\begin{equation}
T_{\mu\nu}=-\frac{2}{e^{2}}\text{Tr}\biggl(F_{\mu \alpha }{F_{\nu}}^{\alpha} -%
\frac{1}{4}g_{\mu \nu }F_{\alpha \beta }F^{\alpha \beta }\biggl)-\frac{1}{2} 
\text{Tr}\biggl(D_{\mu }\varphi D_{\nu }\varphi -\frac{1}{2}g_{\mu \nu
}D_{\alpha }\varphi D^{\alpha }\varphi \biggl) \ .
\end{equation}
In order to construct analytical solutions of the Yang-Mills-Higgs theory in 
$(3+1)$-dimensions we will use as a starting point the same ansatz for the $%
U $ field and the connection $A_{\mu}$ introduced for the case without the
Higgs contribution, namely Eqs. \eqref{ans2} and \eqref{ans1}. Now, for the
Higgs field we must consider the following general form 
\begin{equation}  \label{varphi}
\varphi = \sum_{j=1}^{3} f_j(r) h^j(t,\phi) \mathbf{t}_j \ ,
\end{equation}
where $f_j$ and $h_j$ are functions to be found.

A good choice for the functions introduced above that allows to reduce
significantly the field equations of the Yang-Mills-Higgs system is the
following 
\begin{gather}  \label{hs}
h_1(t,\phi)=\frac{a}{b} h(t,\phi) \ , \quad h_3(t,\phi)=a \cot
\left(H(t,\phi)\right)\frac{h(t,\phi)}{\lambda(t,\phi)} \ , \quad
\lambda_3=1 \ , \\
f_1(r)=b \cos(qr)f_3(r) \ , \quad f_2(r)=a \sin(q r) f_3(r) \ , \quad
f_3(r)=f_0 \,r \ ,  \label{fs}
\end{gather}
where we have defined 
\begin{equation*}
h_2(t,\phi):= h(t,\phi) \ , \qquad \lambda_1(t,\phi)=\lambda_2(t,\phi)
:=\lambda(t,\phi) \ ,
\end{equation*}
being $a$, $b$, and $f_0$ arbitrary constants.

In fact, it is direct to check that Eqs.~\eqref{varphi}, \eqref{hs} and %
\eqref{fs}, together with Eqs. \eqref{ans2} and \eqref{ans1}, reduce the
complete set of Yang-Mills-Higgs equations to the following decoupled
partial differential equations 
\begin{equation*}
\Box H=\left( \frac{\partial ^{2}}{\partial t^{2}}-\frac{1}{L_{\phi }^{2}}%
\frac{\partial ^{2}}{\partial \phi ^{2}}\right) H=0\,,\qquad \Box h=\left( 
\frac{\partial ^{2}}{\partial t^{2}}-\frac{1}{L_{\phi }^{2}}\frac{\partial
^{2}}{\partial \phi ^{2}}\right) h=0\,,\qquad \Box \lambda =\left( \frac{%
\partial ^{2}}{\partial t^{2}}-\frac{1}{L_{\phi }^{2}}\frac{\partial ^{2}}{%
\partial \phi ^{2}}\right) \lambda =0\,,
\end{equation*}%
together with 
\begin{eqnarray}
\left( \frac{\partial H}{\partial t}\right) ^{2}-\frac{1}{L_{\phi }^{2}}%
\left( \frac{\partial H}{\partial \phi }\right) ^{2} &=&\left( \frac{%
\partial H}{\partial t}-\frac{1}{L_{\phi }}\frac{\partial H}{\partial \phi }%
\right) \left( \frac{\partial H}{\partial t}+\frac{1}{L_{\phi }}\frac{%
\partial H}{\partial \phi }\right) =0\,, \\
\left( \frac{\partial h}{\partial t}\right) ^{2}-\frac{1}{L_{\phi }^{2}}%
\left( \frac{\partial h}{\partial \phi }\right) ^{2} &=&\left( \frac{%
\partial h}{\partial t}-\frac{1}{L_{\phi }}\frac{\partial h}{\partial \phi }%
\right) \left( \frac{\partial h}{\partial t}+\frac{1}{L_{\phi }}\frac{%
\partial h}{\partial \phi }\right) =0\,, \\
\left( \frac{\partial \lambda }{\partial t}\right) ^{2}-\frac{1}{L_{\phi
}^{2}}\left( \frac{\partial \lambda }{\partial \phi }\right) ^{2} &=&\left( 
\frac{\partial \lambda }{\partial t}-\frac{1}{L_{\phi }}\frac{\partial
\lambda }{\partial \phi }\right) \left( \frac{\partial \lambda }{\partial t}+%
\frac{1}{L_{\phi }}\frac{\partial \lambda }{\partial \phi }\right) =0\,.
\end{eqnarray}%
Additionally, from the Yang-Mills equations the following first order
non-linear equation emerges 
\begin{equation}
\frac{\partial \lambda }{\partial t}+\tan (H)\lambda \left( 1-\lambda
^{2}\right) \frac{\partial H}{\partial t}=0~~~\Rightarrow ~~~\lambda =\pm 
\frac{\cos (H)}{\sqrt{\exp (2\lambda _{0})+\cos ^{2}(H)}}\ ,
\label{lambdasol}
\end{equation}%
that fixes the function $\lambda $ in terms of $H$ (here $\lambda _{0}$ is
constant). Hence, the constraint here above reduces the number of chiral
modes to two.

Summarizing, with the ansatz presented above, the complete set of field
equations of the Yang-Mills-Higgs theory has been reduced to the field
equations of three chiral massless scalar fields in $(1+1)$-dimensions plus
a non-linear constraint between two of them. Consequently, these families of
exact solutions with non-vanishing topological charge are labelled by two
integers ($p$ and $q$, which determine the topological charge in Eq.~(\ref%
{aux2})) and two chiral massless fields in $(1+1)$-dimensions (namely $H$
and $h$), since $\lambda $ depends on $H$ as in Eq. \eqref{lambdasol}. Quite
interestingly, the inclusion of the Higgs field leads to two-dimensional
chiral massless modes (instead of massless modes).

The energy density $T_{00}^{(1)}$ of the above solutions takes the form %
\begin{align}\nonumber
T_{00}^{(1)}\ =& \frac{(1+e^{2\lambda_0})}{2}\biggl(\csc^2(H) \biggl[a^2 f_0^2 r^2h^2\cot^2(H) ~+~\frac{e^{4\lambda_0}p^2\sin^4(H)}{e^2L_\theta^2(e^{2\lambda_0}+\cos^2(H))^3}   \biggl] \biggl((\partial_t H)^2+\frac{1}{L_\phi^2}(\partial_\phi H)^2\biggl)\\
& +~a^2f_0^2 r^2\csc^2(H) \biggl((\partial_t h)^2+\frac{1}{L_\phi^2}(\partial_\phi h)^2\biggl)~+~\frac{a^2 f_0^2}{ L_r^2} h\csc^2(H) \biggl[h-4L_r^2 r^2\cot(H)\partial_t H\partial_t h\biggl] \biggl) \ ,  \label{tmunuymh}
\end{align}%
where $\partial _{t}$ and $\partial _{\phi}$ stand for derivative, respectively, with respect to $t$ and $\phi$ and the field equations have been used in order to reduce the last term. Here it is
worth to note the following fact: at a first glance, because the ansatz
reduces the complete set of Yang-Mills-Higgs field equations to a set of
linear decoupled equations (one for $H$ and one for $h$), one could suspect
that perhaps the above configurations of Yang-Mils-Higgs theory are, after
all, gauge equivalent to Abelian non-interacting configurations. However, if
this would be the case, then the energy-density (which is gauge-invariant)
should also be the energy density of two decoupled chiral massless modes
(which is quadratic in the fields, satisfies linear equations and only contains kinetic terms of the chiral fields). In the
present case, the above expression for the energy density clearly manifests
non-linear interactions between the two main degrees of freedom $H$ and $h$. 

On the other hand, the CS density becomes 
\begin{equation}
\rho _{\text{CS}}=-\frac{1}{16\pi ^{2}}pq\sin (H)\frac{\partial H}{\partial
\phi }\ .
\end{equation}%
Integrating in the ranges defined in Eq.~\eqref{ranges1}, the topological
charge turns out to be $Q_{\text{CS}}=pq$, where we have used the following
boundary conditions 
\begin{equation*}
H(t,\phi =\pi )=0\ ,\qquad H(t,\phi =0)=\pi \ .
\end{equation*}


\section{Non-linear sigma model in $(3+1)$-dimensions}


Here and in the following sections, we will discuss the NLSM and the Skyrme
model in $(3+1)$-dimensions in the $SU(2)$ case, which is more relevant than
Yang-Mills-Higgs theory as far as the low energy phase diagram of QCD.
Hence, the primary variable will be an $SU(2)$-valued scalar field $U$. We
will analyze how one can construct in these non-integrable theories an
infinite-dimensional family of exact solutions labelled by two integers, as
well as by a free massless scalar field in two dimensions keeping alive the
topological charge, which (in this case as well) can be interpreted as the
Baryonic charge. The key technical point is to find a suitable ansatz which,
on the one hand, depends on all the four space-time coordinates (for the
topological density to be non-vanishing) and, at the same time, reduces the
field equations to the field equations of a free massless scalar field in
two dimensions. The high physical interest in the NLSM can be quickly
explained, considering its many relevant physical applications. In
particular, as far as the present paper is concerned, the model is related
to the low energy limit of QCD and Pion's dynamics (see \cite%
{shifman1,shifman2} and references therein). Thus, the current approach can
provide an infinite family of topologically non-trivial solutions allowing
the explicit computation of critical physical quantities (which would be
impossible to obtain from perturbation theory). In fact, in many situations
of physical interest (especially at finite Baryon density), both
perturbation theory and even the powerful tools of LQCD may fail (see \cite%
{[7], [8],[9]} and references therein).

The action of the $SU(2)$-NLSM in $(3+1)$-dimensions is 
\begin{equation}
I[U]=\frac{K}{4}\,\int d^{4}x\sqrt{-g}\,\mathrm{Tr}\left( R^{\mu }R_{\mu
}\right) \ ,  \label{INLSM1}
\end{equation}%
where $K$ is the coupling constant of the NLSM and $R_{\mu }$ has been
defined in Eq.~\eqref{R}. It is worth emphasizing that the NLSM only
possesses global symmetry and is not classically conformal invariant in $%
(3+1)$-dimensions (unlike Yang-Mills theory). Nevertheless, despite the
enormous differences between these two theories, an approach similar to the
one described in the previous section also works in the present case. The
field equations obtained varying the action in Eq.~\eqref{INLSM1} with
respect to the $U$ field are 
\begin{equation}
\nabla _{\mu }R^{\mu }=0\,,  \label{NLSM}
\end{equation}%
and the energy-momentum tensor of the model is 
\begin{equation}
T_{\mu \nu }=-\frac{K}{2}\mathrm{Tr}\left[ R_{\mu }R_{\nu }-\frac{1}{2}%
g_{\mu \nu }R^{\alpha }R_{\alpha }\right] \ .  \label{NLSMtmunu}
\end{equation}

\subsection{CFT in two dimensions from the NLSM}


We will use the metric in Eq.~\eqref{Box4} whose ranges for the coordinates
can be determined in a similar way as in Eq.~\eqref{ranges1} (where the
theory of Euler angles came into play). Let us define the following $U(x)\in
SU(2)$ 
\begin{equation}
U=\exp \left( p\,\theta \frac{\mathbf{t}_{3}}{2}\right) \,\exp \left( r\frac{%
\mathbf{t}_{2}}{4}\right) \,\exp \left( F\left( t,\phi \right) \frac{\mathbf{%
t}_{3}}{2}\right) \,,  \label{NLSM1}
\end{equation}%
where $p$ is a non-vanishing integer (there will be one more restriction to
be discussed later on). The theory of Euler angles for $SU(N)$ \cite%
{euler1,euler2,euler3} tells that the range of $\theta $ (appearing in the
left factor of the decomposition in Eq.~\eqref{NLSM1}) and the range of $r$
(appearing in the central factor of the decomposition in Eq.~\eqref{NLSM1})
must be 
\begin{equation}
0\leq \theta \leq \pi \,,\qquad 0\leq r\leq 2\pi \ .\   \label{NLSMrange1}
\end{equation}%
One can also consider the range of the coordinate $\phi $ as 
\begin{equation}
0\leq \phi \leq 2\pi \ .  \label{NLSMrange2}
\end{equation}%
As far as the exponent in the right factor (namely $F\left( t,\phi \right) $%
) is concerned, there are again two options. If the field $F\left( t,\phi
\right) $ satisfies periodic boundary conditions then the topological charge
of the $SU(2)$-valued scalar field $U$ vanishes (although the topological
density in Eq.~\eqref{rhoB} can still be non-trivial). The other boundary
condition for $F\left( t,\phi \right) $ arises naturally taking into account
two facts. First of all, one has to require that physical observables (built
from traces of product of the $SU(2)$-valued field $U$ and its derivatives)
such as the energy-momentum tensor should be periodic in $\phi $ and this
requirement \textit{does not imply} that $F\left( t,\phi \right) $ itself is
periodic. Secondly, $F\left( t,\phi \right) $ appears in the right factor of
the Euler angles decomposition of an $SU(2)$ element (see, for instance,
Refs. \cite{euler1,euler2,euler3}) 
\begin{equation}
F\left( t,\phi =0\right) -F\left( t,\phi =2\pi \right) =\pm\, 8\,q\,\pi \ ,
\label{BCNLSM3}
\end{equation}%
where $q$ is a non-vanishing integer. The option here above ensures that the 
$SU(2)$ valued element $U$ defined in Eqs.~\eqref{NLSM1} and \eqref{BCNLSM3}
wraps an integer number of times around the group manifold of $SU(2)$ (in
other words, $U$ has a non-vanishing winding number). In this case, the
topological charge and the topological density associated with $U$ will be
non-trivial. Also, in the present section, the term \textquotedblleft
topologically non-trivial'' refers to configurations with $\rho _{\text{B}%
}\neq 0$: the reason is that configurations with vanishing total Baryonic
charge but non-vanishing $\rho _{\text{B}}$ still describe non-trivial
interacting configurations with both regions having positive and negative
charge densities.

It is an astounding and powerful result (due to all the analytic
non-perturbative tools that will become available) that, despite the
non-integrable character of the NLSM in $(3+1)$-dimensions, the complete set
of NLSM field equations in Eq.~\eqref{NLSM} corresponding to the ansatz in
Eq.~\eqref{NLSM1} reduce to the field equation of a free massless scalar
field in two dimensions keeping alive the topological charge density 
\begin{equation}
\left( \frac{\partial ^{2}}{\partial t^{2}}-\frac{1}{L_{\phi}^{2}}\frac{%
\partial ^{2}}{\partial \phi ^{2}}\right) F(t,\phi)=0\ .  \label{NLSMequ}
\end{equation}
It is worth emphasizing that $F$ represents a Goldstone mode associated to
the phase of the two charged Pions. In other words, if one would associate a
complex wave function to the two charged Pions (keeping out of such wave
function the neutral Pion) then the scalar field $F$ would be the phase of
the wave function.

\subsection{Topological charge and energy density}


With the ansatz in Eq.~\eqref{NLSM1} the topological density and topological
charge, respectively, read 
\begin{eqnarray}
\rho _{\text{B}} &=&-\frac{p}{32\pi ^{2}}\sin \left( \frac{r}{2}\right)\frac{%
\partial F}{\partial \phi }\ ,  \label{topdensNLSM} \\
B &=&-\frac{p}{8\pi }\left[ F\left( t,\phi =2\pi \right) -F\left( t,\phi
=0\right) \right] =\pm\, pq\ .  \label{topchargeNLSM}
\end{eqnarray}%
It is worth noting that the topological charge density in Eq.~%
\eqref{topdensNLSM} has a non-trivial profile depending both on $r$ and on\ $%
\phi $. The maximum of $\rho _{\text{B}}$ are located at $r=\pi $ and at the
values of $\phi $ such that $\partial F/\partial \phi $ is maximum: in three
spatial dimensions these two conditions identify a line. As long as $%
\partial F/\partial \phi \neq 0$, the topological density is non-zero. Note
that the topological density is a linear function of $F\left( t,\phi \right) 
$, different from the Yang-Mills case presented in the previous section.

The energy density reads 
\begin{equation}
T_{00}^{\sigma}=\frac{K}{8}\biggl[\frac{1}{4}\biggl(\frac{1}{L_{r}^{2}}+%
\frac{4p^{2}}{L_{\theta }^{2}}\biggl)+ \biggl(\frac{\partial F}{\partial t}%
\biggl)^{2}+\frac{1}{L_{\phi }^{2}}\biggl(\frac{\partial F}{\partial \phi }%
\biggl)^{2} \biggl]\,,  \label{T00a}
\end{equation}%
then the total energy is given by 
\begin{eqnarray}
E^{\sigma}& =& \int \sqrt{-g } dr d\theta d\phi \,T_{00}^{\sigma}\,,  \notag
\\
& =& \Gamma^{\sigma}+\Psi^{\sigma}\int_{0}^{2\pi}\hspace{-0.3cm}d\phi\left(
\left( \frac{\partial F}{\partial t} \right)^2 +\frac{1}{L_{\phi}^{2}}
\left( \frac{\partial F}{\partial \phi} \right)^2 \right) \,,
\end{eqnarray}
where 
\begin{equation}  \label{energy1}
\Gamma^{\sigma} = \frac{K \pi^3 L_{\phi} }{8 L_{r} L_{\theta}}\,\left(
L_{\theta}^{2}+ 4p^2L_r^2\right)\,,~~~~~~\Psi^{\sigma} = \frac{K \pi^2 L_r
L_{\theta} L_{\phi}}{4 }\,.
\end{equation}
On the other hand, the on-shell action becomes 
\begin{equation}
I^{\sigma}_{\text{on-shell}}[F]=-\frac{K}{8}\int \sqrt{-g}drd\theta d\phi %
\biggl[\frac{1}{4}\biggl(\frac{1}{L_{r}^{2}}+\frac{4p^{2}}{L_{\theta }^{2}}%
\biggl)- \biggl(\frac{\partial F}{\partial t}\biggl)^{2}+\frac{1}{L_{\phi
}^{2}}\biggl(\frac{\partial F}{\partial \phi }\biggl)^{2} \biggl]\ .
\end{equation}%
It is important to note that the energy does not grow linearly with the
topological charge, as can be seen from Eqs.~\eqref{topchargeNLSM} and %
\eqref{T00a}. This fact indicates that these solutions describe interacting
systems (as otherwise, the energy would be linear in the topological charge).

\subsection{Conformal field theory and some semiclassical considerations}


The usual mode expansion of the solutions of Eq.~\eqref{NLSMequ} is of
course the same as in the previous section in Eqs.~\eqref{general1} and (\ref%
{general2}), where $F_{+}$ refers to the left movers and $F_{-}$ to the
right movers ($v_{\pm }$ and $\phi _{0}^{\pm }$\ being integration constants
which must satisfy three constraints which will be discussed below). Hence,
the most general topologically non-trivial configuration of the present
sector arises replacing, $F=F_{+}+F_{-}$, in Eqs. (\ref{general1}) and (\ref%
{general2}) into Eq. (\ref{NLSM1}).

Also, in the present case, the most natural choice corresponds to take $%
a_{n}^{\pm }$ and $b_{n}^{\pm }$ in such a way that 
\begin{equation*}
\widetilde{F}(t,\phi =0)=\widetilde{F}(t,\phi =2\pi )=0\ ,
\end{equation*}%
where $\widetilde{F}(t,\phi )$ is the part of $F=F_{+}+F_{-}$ coming from
the sum over the integers $n$ in Eqs. (\ref{general1}) and (\ref{general2}).
Therefore $B$ in Eq. (\ref{topchargeNLSM}) is non-zero when $v_{+}-v_{-}\neq
0$. Also, $v_{\pm }$ in Eqs. (\ref{general1}) and (\ref{general2}) must be
chosen as%
\begin{equation}
F\left( t,\phi =0\right) =\phi _{0}^{+}+\phi _{0}^{-}+\left(
v_{+}+v_{-}\right) \frac{t}{L_{\phi}}~~~~ \Rightarrow ~~~v_{+}+v_{-}=0\ ,
\label{NLSMchoice1}
\end{equation}%
\begin{equation}
F\left( t,\phi =2 \pi \right) =\phi _{0}^{+}+\phi _{0}^{-}+\left(
v_{+}-v_{-}\right) 2 \pi ~~~~ \Rightarrow ~~~v_{+}-v_{-}=4q\ .
\label{NLSMchoice2}
\end{equation}%
Unlike what happens in the Yang-Mills case, here there is no constraint on $%
\phi _{0}^{+}+\phi _{0}^{-}$. Hence, the topological charge is%
\begin{equation*}
B=p\, q\ .
\end{equation*}%
At the classical level, this is the most straightforward possible choice of
boundary conditions since it allows to identify the terms in the expansion
modes responsible for the topological charge and which are not. However,
plenty of different options will be discussed in forthcoming papers.

Also, in the present case, the semi-classical quantization of these
configurations corresponds to the quantization of the free massless scalar
field $F\left( t,\phi \right) $ with the boundary conditions described above
to have a non-vanishing topological charge. However, as discussed in the
previous sections, some interesting differences exist.

First, in Eqs. (\ref{general1}) and (\ref{general2}) any term in
the expansion corresponds to an exact solution of the $(3+1)$-dimensional
NLSM field equations and not just to a solution of the linearized field
equations. Therefore, the Bosonic quantum operators $\alpha _{n}^{+}$, $%
\left( \alpha _{m}^{+}\right) ^{\dag }$ and $\alpha _{n}^{-}$, $\left(
\alpha _{m}^{-}\right) ^{\dag }$ (which are annihilation and creation
operators for the left and right movers, satisfying the obvious commutation
relations, see \cite{GSW}) are quantum operators which create exact
solutions of the semiclassical NLSM field equations.

Second, the constant terms $\phi _{0}^{\pm }$\ as well as the
linear terms in $t$ and $\phi $\ play an important role as these are
associated to classical solutions which carry the topological charge (while
the modes satisfying periodic boundary conditions do not contribute to the
topological charge). Thus, depending on whether $B$ is odd or even, one
should quantize the modes associated to the linear terms in the expansion of 
$F$ as Fermionic or Bosonic. Hence, when $B$ is odd, $F$ has a component
which should be considered as an emergent Fermionic field.


\section{The Skyrme model in $(3+1)$-dimensions}


A very natural question is: \textit{does the Skyrme term spoil the
remarkable relation discussed in the previous section between the simplest two-dimensional CFT and a non-integrable theory in $(3+1)$-dimensions at finite Baryon density in topologically non-trivial sectors?} The importance of the Skyrme model lies in the fact that the NLSM in flat space-time does not admit static topologically non-trivial soliton solutions with finite energy, known as Derrick's scale argument \cite{Derrick}. The Skyrme term is introduced to get around this problem and stabilize the
soliton (Skyrmion).

The obvious physical relevance of finite density effects arises from the
difficulties in providing cold and dense nuclear matter as a function of
baryon number density with a good analytic understanding. The
non-perturbative nature of low energy QCD prevents (the very complex and
interesting structure of) its phase diagram from being described in detail
(see \cite{R1,R2,newd3,newd4,newd5,newd6} and references therein): this is
the reason why researchers in this area mainly use numerical and lattice
approaches. In particular, a very intriguing part in the QCD phase diagram,
which appears at finite baryon density,\footnote{%
See \cite{[80],[81],[82],[83],[84],[85],Schmitt1,Schmitt2} and references
therein, for the construction of non-homogeneous condensates at finite
density in chiral perturbation theory.} is related to the appearance of
ordered structures (similar to the Larkin--Ovchinnikov--Fulde--Ferrell phase 
\cite{LOFF1yLOFF2}). These ordered structures at finite density are, by now,
a well-established feature (see, for instance, \cite%
{newd7,newd13y16,newd20toN3}, and references therein). These are just some
of the reasons why it is mandatory to shed more light on these issues with
theoretical tools, as often even the numerical approaches are not effective
with high topological charges.

Here we will show that the Skyrme term discloses a remarkable phenomenon:
namely, the present construction still works (with precisely the same
ansatz) but now, when the Skyrme coupling is non-zero, instead of a
two-dimensional CFT, one gets a two-dimensional chiral CFT: namely, either
left or right movers must be eliminated. This new result is likely to be
related to the fact that the Skyrme model includes the effects of the low
energy limit of QCD so that the Skyrme model knows, somehow, about chiral
symmetry breaking.

The Skyrme action is given by 
\begin{equation*}
I[U]=\frac{K}{4}\,\int d^{4}x\sqrt{-g}\,\text{Tr}\biggl(R_{\mu }R^{\mu }+%
\frac{\lambda }{8}[R_{\mu },R_{\nu }][R^{\mu },R^{\nu }]\biggl)\ ,
\end{equation*}%
where $K$ and $\lambda $ are positive coupling constants.\footnote{The parameters $K$ and $\lambda$
are related to the meson decay coupling constant $F_{\pi}$ and the Skyrme coupling $e$
via $F_{\pi}=2 \sqrt{K}$ and $K \lambda e^{2}=1$, where $F_{\pi}=141 \text{MeV}$ and $e=5.45$.}
The field
equations of the model are obtained varying the last action with respect to
the $U$ field, we get 
\begin{equation}
\nabla ^{\mu }\biggl(R_{\mu }+\frac{\lambda }{4}[R^{\nu },[R_{\mu },R_{\nu
}]]\biggl)\ =\ 0\ ,  \label{Eqs}
\end{equation}%
being these three non-linear coupled second-order partial differential
equations.

The energy-momentum tensor reads 
\begin{equation}
T_{\mu \nu }\ =\ -\frac{K}{2}\text{Tr}\biggl(R_{\mu }R_{\nu }-\frac{1}{2}%
g_{\mu \nu }R_{\alpha }R^{\alpha }+\frac{\lambda }{4}\left(g^{\alpha \beta
}[R_{\mu },R_{\alpha }] [R_{\nu },R_{\beta }] -\frac{1}{4}g_{\mu \nu
}[R_{\alpha },R_{\beta }] [R^{\alpha },R^{\beta }] \right) \biggl)\ .
\label{Tmunu}
\end{equation}%
The topological density and charge are defined in Eqs.~\eqref{rhoB} and %
\eqref{charges}. Now, we will study two types of analytical configurations
that will lead to a chiral CFT. The description of the box is based on the
metric given in Eqs.~\eqref{Box4}, \eqref{NLSMrange1} and \eqref{NLSMrange2}.

\subsection{Chiral conformal field theory from the Skyrme model. Type-I:
Euler ansatz for the lasagna phase}


We will consider, once again, the matter field ansatz in Eq.~\eqref{NLSM1}.
When one plugs Eq. (\ref{NLSM1}) into the Skyrme equations in Eq.~\eqref{Eqs}%
, the field equations reduce to%
\begin{gather}
\left( \frac{\partial ^{2}}{\partial t^{2}}-\frac{1}{L_{\phi }^{2}}\frac{%
\partial ^{2}}{\partial \phi ^{2}}\right) F=0\ ,  \label{skyrmetypeI1} \\
\left( \frac{\partial F}{\partial t}\right) ^{2}-\frac{1}{L_{\phi }^{2}}%
\left( \frac{\partial F}{\partial \phi }\right) ^{2}=\left( \frac{\partial F%
}{\partial t}-\frac{1}{L_{\phi }}\frac{\partial F}{\partial \phi }\right)
\left( \frac{\partial F}{\partial t}+\frac{1}{L_{\phi }}\frac{\partial F}{%
\partial \phi }\right) =0\ .  \label{skyrmetypeI2}
\end{gather}%
As in the NLSM and Yang-Mills cases, the first equation describes the
simplest Bosonic CFT in two dimensions. Thus, from Eq.~\eqref{skyrmetypeI1} $%
F=F_{+}+F_{-}$ (where $F_{\pm }$ represent the contributions of the left and
right movers). However, Eq.~\eqref{skyrmetypeI2} can be satisfied only by
killing either $F_{+}$ or $F_{-}$. Hence, we still get a two-dimensional
CFT, but this time it is a chiral CFT. Once again, this result is a huge
analytic achievement as the field equations have been reduced exactly,
keeping alive the topological density, to the field equations of a free
massless chiral scalar field in $(1+1)$-dimensions. Also, the topological
charge is the same as in the NLSM case defined in Eq. \eqref{topchargeNLSM}.

\begin{figure}[hbtp]
\centering
\includegraphics[scale=.6]{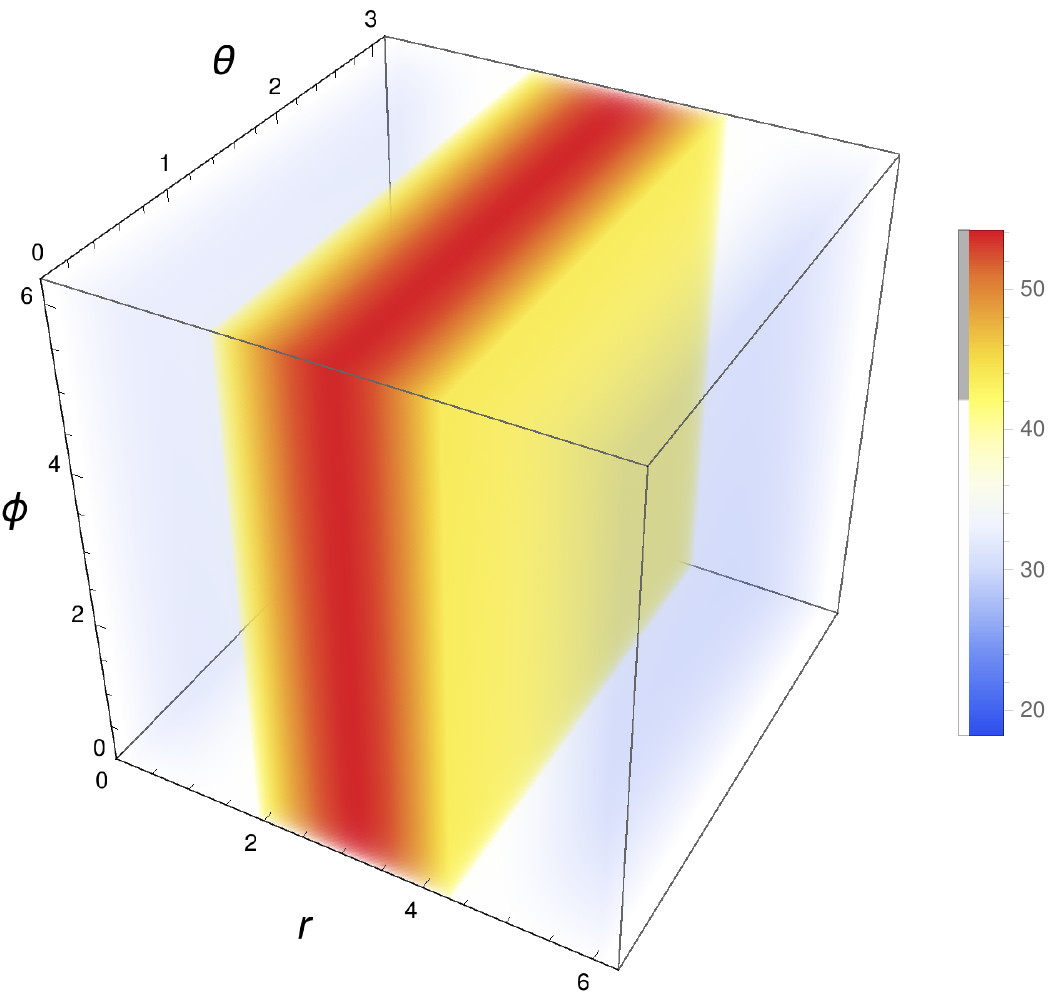}\quad
\includegraphics[scale=.6]{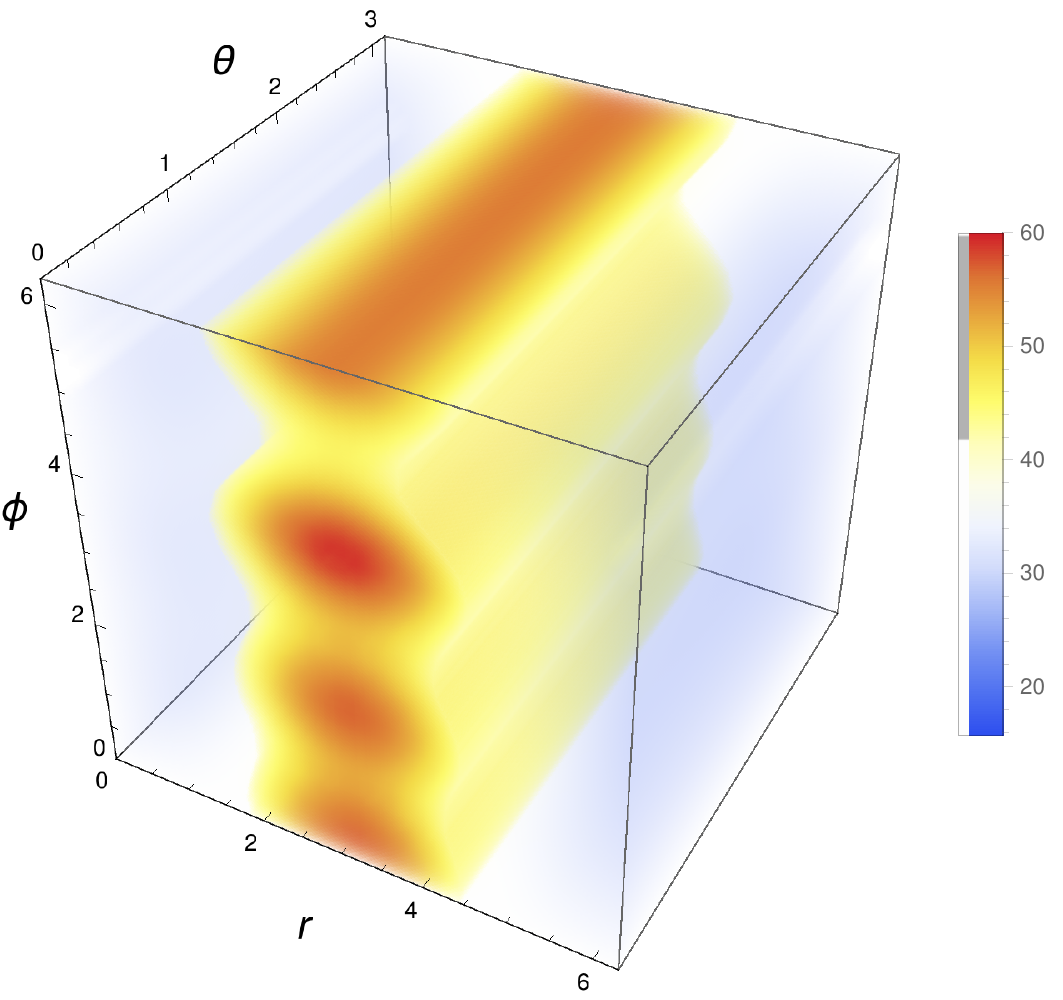}
\caption{Energy density with and without modulation of nuclear lasagna configurations with Baryonic charge $B=6$. For both cases we have set $K=\lambda=L_r=L_\theta=L_\phi=1$, $p=3$, $q=2$ and $\phi_0=0$. Left: nuclear lasagna without modulation where $a_i=b_i=0$. Right: snapshot at t=0 of nuclear lasagna with a modulation in the $\phi$ direction where the non-null modulation coefficients were set as $a_1=-a_3=b_1=b_2=0.1$.}\label{fig:Lasagna}
\end{figure}

In this case, the energy density is given by 
\begin{align}
T_{00}^{(2)}=\ & \frac{K}{8}\biggl[\frac{1}{4}\biggl(\frac{1}{L_{r}^{2}}+%
\frac{4p^{2}}{L_{\theta }^{2}}\biggl)+\left( \frac{\partial F}{\partial t}%
\right) ^{2}+\frac{1}{L_{\phi }^{2}}\left( \frac{\partial F}{\partial \phi }%
\right) ^{2}\biggl]  \notag \\
& +~\frac{K\lambda }{32L_{r}^{2}L_{\theta }^{2}}\biggl[\frac{p^{2}}{4}%
+\left( \frac{1}{4}L_{\theta }^{2}+p^{2}L_{r}^{2}\sin ^{2}\left( \frac{r}{2}%
\right) \right) \biggl(\left( \frac{\partial F}{\partial t}\right) ^{2}+%
\frac{1}{L_{\phi }^{2}}\left( \frac{\partial F}{\partial \phi }\right) ^{2}%
\biggl)\biggl]\ ,  \label{tmunuskl}
\end{align}%
so that, the expression for the energy becomes 
\begin{eqnarray}
E^{(2)} &=&\int \sqrt{-g}\,drd\theta d\phi \,T_{00}^{(2)}  \notag \\
&=&\Gamma ^{(2)}+\Psi ^{(2)}\int_{0}^{2\pi }\hspace{-0.3cm}d\phi \left(
\left( \frac{\partial F}{\partial t}\right) ^{2}+\frac{1}{L_{\phi }^{2}}%
\left( \frac{\partial F}{\partial \phi }\right) ^{2}\right) \,,  \label{E2}
\end{eqnarray}%
where 
\begin{equation}
\Gamma ^{(2)}=\frac{K\pi ^{3}L_{\phi }}{32L_{r}L_{\theta }}\,\left(
4L_{\theta }^{2}+p^{2}(\lambda +16L_{r}^{2})\right) \,,~~~~~~\Psi ^{(2)}=%
\frac{K\pi L_{\phi }}{64L_{r}L_{\theta }}\,\left( \pi L_{\theta}^{2}(\lambda+16
L_{r}^{2})+8p^2 \lambda L_r^2 \right) \,.  \label{energy2}
\end{equation}
As in the NLSM, the energy does not grow linearly with the topological
charge, implying the presence of interactions between particles. Also, the
topological charge density is linear in $F(t,\phi )$ instead of non-linear,
as in the Yang-Mills case presented in the previous sections. These
configurations describe modulated nuclear lasagna layers in which the
periodic part in the mode expansion of the field $F(t,\phi )$ (which does
not carry topological charge) represents the modulations in the $\phi $%
-direction, while the linear part is responsible for the \textquotedblleft
bare lasagna"; namely, the lasagna without modulations which have been
analyzed in \cite{Sergio1} and \cite{Sergio2}). Figure \ref{fig:Lasagna} shows the energy
density of two lasagna-type configurations, one with modulation and the
other without modulation. It is worth emphasizing that it is also necessary to introduce a cut-off in the computation of the (semi)classical partition function because the
Skyrme theory is an effective low-energy model. We will detail this point in
the next section.

\subsection{Chiral conformal field theory from the Skyrme model. Type-II:
Exponential ansatz for the spaghetti phase}


This time for the $U$ field we adopt the standard (exponential)
parameterization of an element of $SU(2)$, that is 
\begin{equation}
U^{\pm 1}\left( x^{\mu }\right) \, = \, \cos (\alpha )\mathbf{1}_{2}~\pm ~\sin
(\alpha )n^{i}\mathbf{t}_{i}\ ,  \label{U}
\end{equation}%
where 
\begin{gather}
n^{1}=\sin \Theta \cos \Phi ,\quad n^{2}=\sin \Theta \sin \Phi ,\quad
n^{3}=\cos \Theta \ ,  \label{n} \\
\alpha =\alpha (x^{\mu })\ ,\quad \Theta =\Theta (x^{\mu })\ ,\quad \Phi
=\Phi (x^{\mu })\ ,\quad n^{i}n_{i}=1\ .  \notag
\end{gather}%
From Eq.~\eqref{rhoB} it follows that the topological charge density takes
the following general form 
\begin{equation}
\rho _{\text{B}}=-\frac{1}{2\pi ^{2}}\sin ^{2}\alpha \sin \Theta \ d\alpha
\wedge d\Theta \wedge d\Phi \ .  \label{rhoBexplicit}
\end{equation}%
Hence, in order to have topologically non-trivial configurations we must
demand that $d\alpha \wedge d\Theta \wedge d\Phi \ \neq \ 0$. On the other
hand, as we want to construct analytical solutions, it is necessary to have
a good ansatz that significantly reduces the Skyrme field equations.
Considering the approach developed in \cite{gaugsk,LastUS1} lead to the
following 
\begin{gather}
\alpha =\alpha (r)\ ,\quad \Theta =Q \,\theta \ ,\quad \Phi =F\left( t,\phi
\right) \ ,  \label{hedgehog1} \\
Q=2v+1\ ,\quad v\in \mathbb{N}\ .  \notag
\end{gather}%
It is a direct computation to verify that, by replacing the ansatz defined
in Eqs.~\eqref{n} and \eqref{hedgehog1} into the Skyrme field equations, one
gets the following system of equations 
\begin{gather}
\left( \frac{\partial ^{2}}{\partial t^{2}}-\frac{1}{L_{\phi }^{2}}\frac{%
\partial ^{2}}{\partial \phi ^{2}}\right) F=0\ ,  \label{skyrmetypeII1} \\
\left( \frac{\partial F}{\partial t}\right) ^{2}-\frac{1}{L_{\phi }^{2}}%
\left( \frac{\partial F}{\partial \phi }\right) ^{2}=\left( \frac{\partial F%
}{\partial t}-\frac{1}{L_{\phi }}\frac{\partial F}{\partial \phi }\right)
\left( \frac{\partial F}{\partial t}+\frac{1}{L_{\phi }}\frac{\partial F}{%
\partial \phi }\right) =0\ ,  \label{skyrmetypeII2}
\end{gather}%
\begin{equation}
\alpha ^{\prime \prime }-\frac{Q^{2}}{2}\frac{(L_{r}^{2}-\lambda \alpha
^{\prime 2})\sin (2\alpha )}{L_{\theta }^{2}+Q^{2}\lambda \sin ^{2}(\alpha )}%
=0\,.  \label{skyrmetypeII3}
\end{equation}%
Once again, the ansatz in Eqs.~\eqref{n} and \eqref{hedgehog1} discloses
many remarkable simplifications. Not only the equation for $\alpha $
decouples from $F$ (when $F$ satisfies Eqs. (\ref{skyrmetypeII1}) and (\ref%
{skyrmetypeII2})) but it can be also reduced to a simple quadrature 
\begin{equation}
\frac{d\alpha }{dr}=\pm\, \eta (\alpha ,E_{0})\ ,\qquad \eta (\alpha ,E_{0})=%
\biggl[\frac{L_{\theta }^{2}}{L_{\theta }^{2}+Q^{2}\lambda \sin ^{2}(\alpha )%
}\biggl(E_{0}-\frac{Q^{2}}{2}\frac{L_{r}^{2}}{L_{\theta }^{2}}\cos (2\alpha )%
\biggl)\biggl]^{\frac{1}{2}}\,,  \label{alpha0}
\end{equation}%
where $E_{0}$ is an integration constant to be fixed by analyzing the
boundary conditions: 
\begin{equation}
F\left( t,\phi =0\right) -F\left( t,\phi =2\pi \right) =2 p\pi \ ,
\end{equation}%
\textbf{\ }and 
\begin{equation*}
\alpha (2\pi )-\alpha (0)=m\pi \ ,\ m\in 
\mathbb{Z}
\ .
\end{equation*}%
In fact, by integrating Eq. \eqref{alpha0} and considering the above
boundary conditions we get to the following equation for $E_{0}$, 
\begin{equation*}
\pm \, m\int_{0}^{\pi }\frac{1}{\eta (\alpha ,E_{0})}d\alpha =2\pi \ .
\end{equation*}%
From the above condition, it is clear that, for large $m$, the integration
constant $E_{0}$ scales as $m^{2}$
\begin{equation*}
E_{0}=m^{2}\xi _{0}\ , \qquad \xi _{0}>0 \ ,
\end{equation*}%
where $\xi _{0}$ (which can also be interpreted as an integration constant)
does not depend on $m$ for large $m$.

Moreover, in this case, Eq. (\ref{skyrmetypeII1}) describes the simplest
Bosonic CFT in two dimensions. Thus, from Eq. (\ref{skyrmetypeII1}) $%
F=F_{+}+F_{-}$ but, once again, Eq.~\eqref{skyrmetypeII2} can be satisfied
only by killing either $F_{+}$ or $F_{-}$. Thus, as in the last case, we
still get a chiral massless scalar field in $(1+1)$-dimensions. We stress
the very intriguing phenomenon of the appearance of chiral modes without the
presence of any actual edge. These chiral modes are \textquotedblleft
hosted" by the Hadronic tubes\footnote{%
This can be seen as follows: the local maxima of the energy density (see Eq. %
\eqref{tmunus} here below), which coincides with the maximum of the
topological density, is found in the center of the tubes, where $\sin
^{2}(\alpha )\sin ^{2}(Q\theta )=1$. The chiral massless modes have their
support around these points. On the other hand, when $\sin ^{2}(\alpha )\sin
^{2}(Q\theta )=0$ the contribution of the chiral modes to the energy density
vanishes.}: hence these configurations describe modulated nuclear spaghetti
configurations. Indeed, the linear part in the mode expansion of the field $%
F(t,\phi )$ is responsible for the \textquotedblleft bare spaghetti",
namely, the nuclear spaghetti without modulations along the axis which have
been analyzed in \cite{Sergio1} and \cite{Sergio2}. On the other hand, the
periodic part in the mode expansion of the field $F(t,\phi )$ (which does
not carry topological charge) represent the modulations of the tubes in the $%
\phi $-direction.

\begin{figure}[hbtp]
\centering
\includegraphics[scale=.6]{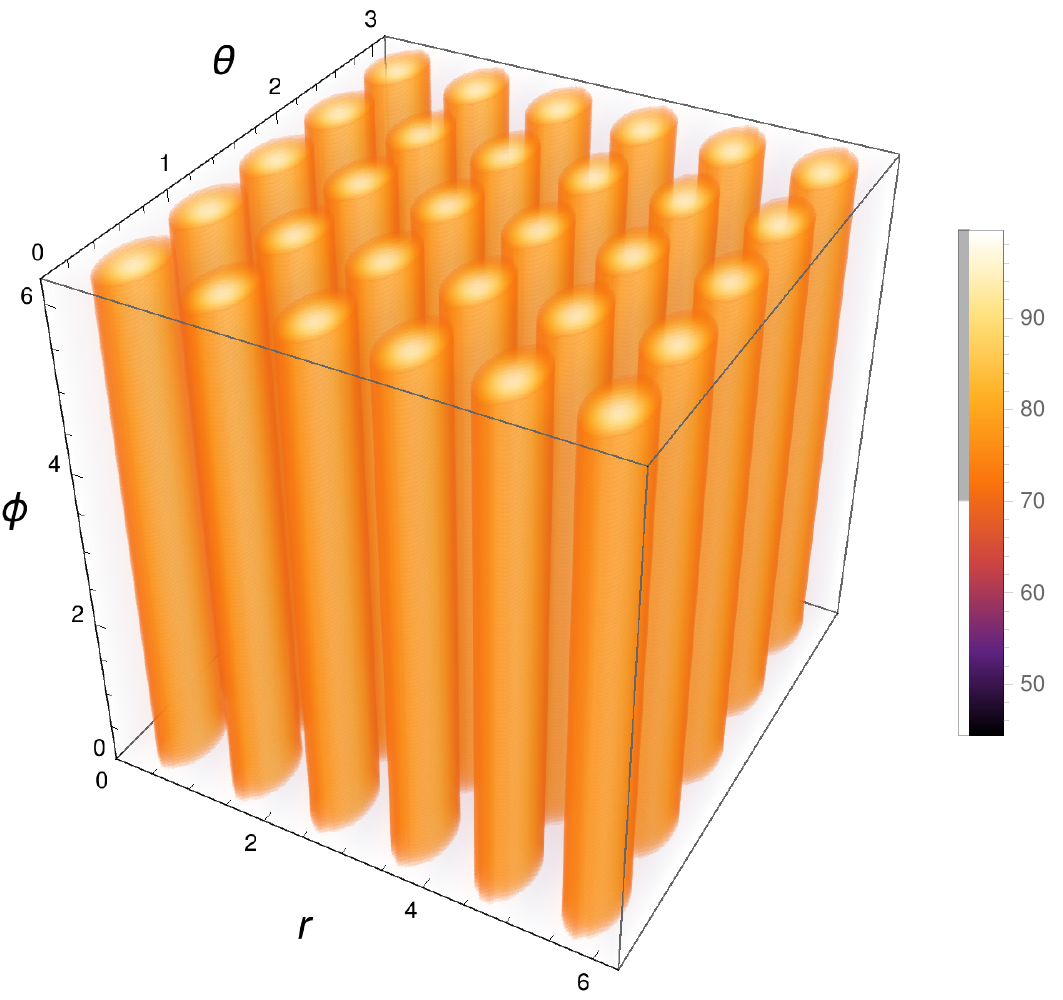}\quad
\includegraphics[scale=.6]{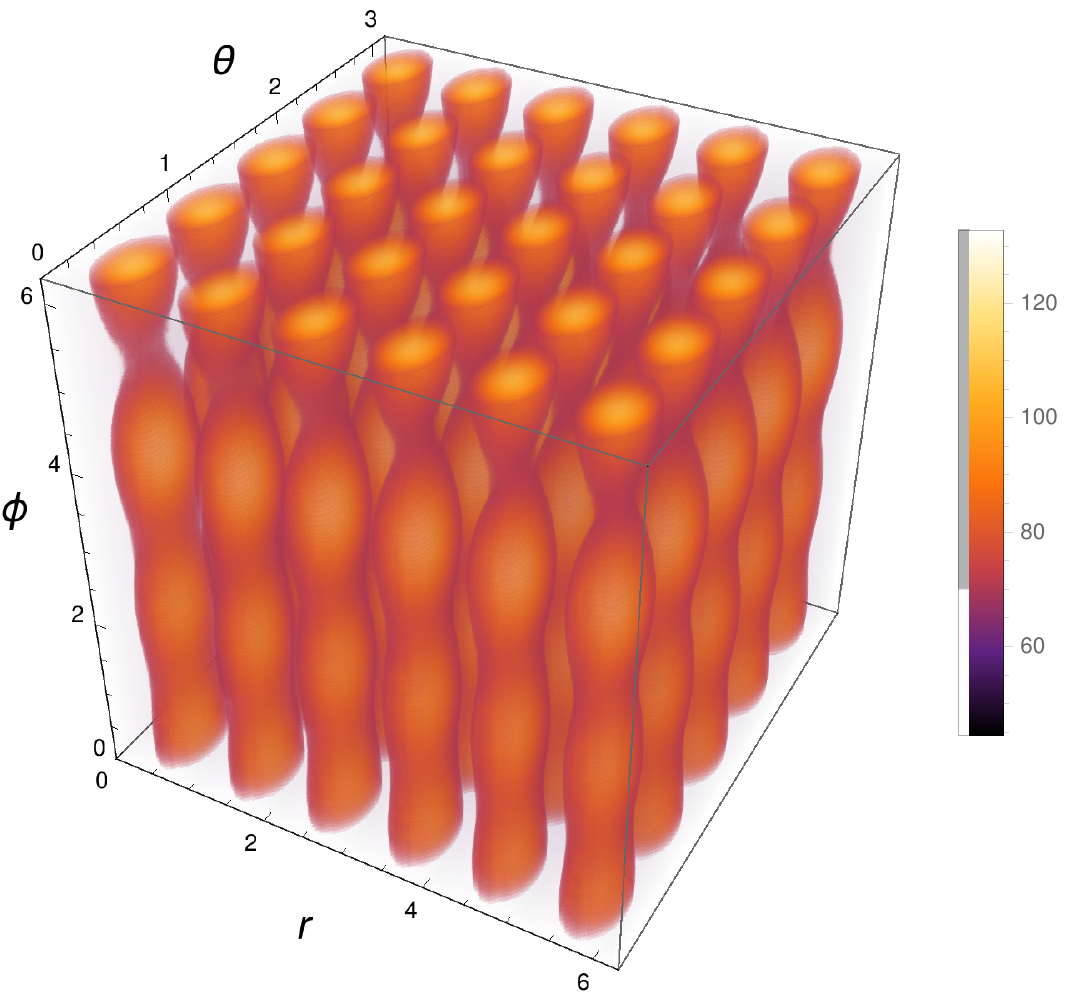}
\caption{Energy density with and without modulation of nuclear spaghetti configurations with Baryonic charge $B=6$. For both cases we have set $K=\lambda=L_r=L_\theta=L_\phi=1$, $p=1$, $m=6$, $q=5$, and $\phi_0=0$. Left: nuclear spaghetti without modulation where $a_i=b_i=0$. Right: snapshot at $t=0$ of nuclear spaghetti with a modulation in the $\phi$-direction where the non-null modulation coefficients were set as $a_1=-a_3=b_1=b_2=0.1$.}\label{fig:Spaghetti}
\end{figure}

The energy density is given by 
\begin{align}
T_{00}^{(3)}\ =\ & \frac{K}{2}\biggl\{\frac{\alpha ^{\prime 2}}{L_{r}^{2}}+%
\biggl[\frac{Q^{2}}{L_{\theta }^{2}}+\biggl(\left( \frac{\partial F}{%
\partial t}\right) ^{2}+\frac{1}{L_{\phi }^{2}}\left( \frac{\partial F}{%
\partial \phi }\right) ^{2}\biggl)\sin ^{2}(Q\theta )\biggl]\sin ^{2}(\alpha
)\biggl\}  \label{tmunus} \\
& +\frac{K\lambda }{2}\biggl\{\frac{Q^{2}}{L_{\theta }^{2}}\sin ^{2}(Q\theta
)\sin ^{2}(\alpha )\biggl(\left( \frac{\partial F}{\partial t}\right) ^{2}+%
\frac{1}{L_{\phi }^{2}}\left( \frac{\partial F}{\partial \phi }\right) ^{2}%
\biggl)+\frac{\alpha ^{\prime 2}}{L_{r}^{2}}\biggl[\frac{Q^{2}}{L_{\theta
}^{2}}+\sin ^{2}(Q\theta )\biggl(\left( \frac{\partial F}{\partial t}\right)
^{2}+\frac{1}{L_{\phi }^{2}}\left( \frac{\partial F}{\partial \phi }\right)
^{2}\biggl)\biggl]\biggl\}\sin ^{2}(\alpha )\,,  \notag
\end{align}%
then, the total energy is given by 
\begin{align}
E^{(3)} &= \int \sqrt{-g}\,drd\theta d\phi \,T_{00}^{(3)}  \,, \notag \\
&= \Gamma ^{(3)}+~\Psi ^{(3)}\,\int_{0}^{2\pi } d\phi \left[ \left( \frac{%
\partial F}{\partial t}\right) ^{2}+\frac{1}{L_{\phi }^{2}}\left( \frac{%
\partial F}{\partial \phi }\right) ^{2}\right] \,,  \label{E3}
\end{align}%
where 
\begin{equation*}
\Gamma ^{(3)}=\frac{m K\pi ^{2}L_{\phi }}{L_{r} L_{\theta }}\int_{0}^{\pi }%
\hspace{-0.2cm}d\alpha \,\Omega (\alpha ,m,Q)\,,\qquad \Psi ^{(3)}=\frac{%
mK\pi L_{\phi }}{4L_{r}L_{\theta }}\,\int_{0}^{\pi }\hspace{-0.2cm}d\alpha \,%
\tilde{\Omega}(\alpha ,m,Q) \ ,
\end{equation*}%
and 
\begin{eqnarray}
\Omega (\alpha ,m,Q) &=&\eta (\alpha ,E_{0})\left( L_{\theta }^{2}+\lambda
Q^{2}\sin^2(\alpha)\right) +\frac{L_{r}^{2}Q^{2}}{\eta (\alpha ,E_{0})}\sin
^{2}(\alpha ) \ ,  \label{omega1} \\
\tilde{\Omega}(\alpha ,m,Q) &=&\eta (\alpha ,E_{0})\lambda L_{\theta
}^{2}\sin ^{2}(\alpha )+\frac{\sin ^{2}(\alpha )}{\eta (\alpha ,E_{0})}%
L_{r}^{2}\left( L_{\theta }^{2}+\lambda Q^{2}\sin^2(\alpha)\right) \,,
\label{omega2}
\end{eqnarray}
while the topological charge density reads 
\begin{equation}
\rho _{\text{B}}=\frac{1}{2\pi ^{2}}\left( \sin ^{2}(\alpha )\alpha ^{\prime
}\right) \left( \sin \left( Q\theta \right) \right) \left( \partial _{\phi
}F\right) dr\wedge d(Q\theta) \wedge d\phi \,.  \label{spaGnocchi}
\end{equation}
Note that the positions of the maximum of $\rho _{\text{B}}$ are located at 
\begin{equation*}
Q\,\theta =\frac{\pi }{2}+N\pi \,,~~~~~ \sin ^{2}(\alpha )=1\ ,
\end{equation*}%
($N$ being an integer) and at the values of $r$ and $\phi $ such that both $%
\sin ^{2}(\alpha )\alpha ^{\prime }$ and $\partial _{\phi }F$ have maximum.
In three spatial dimensions these three conditions identify isolated points,
and the same happens for the energy density of these configurations (as we
mentioned above). Taking into account the boundary conditions satisfied by $%
\alpha $ and $F$ one arrives at the following value of the Baryonic charge:%
\begin{equation*}
B=m\,p\ .
\end{equation*}%
It is important to emphasize that both the energy density and the
topological charge density depend on all the three spatial coordinates: to the best of the authors' knowledge, these are the first analytic examples of solitons crystals in which both the energy density and the Baryon density manifest a genuine three-dimensional behaviour. Figure \ref{fig:Spaghetti} shows the energy density of two spaghetti-type configurations, one with modulation and the other without modulation.

\section{Partition functions}


This section will discuss the semi-classical partition function
associated with some of the families of topologically non-trivial
configurations constructed in the previous sections. The wording ``semi-classical partition functions" in this section refers to the following: all the exact solutions described previously are characterized both by \textit{some discrete labels} (which determine the Baryonic charge) and (for any possible choice of the discrete labels) by a \textit{massless chiral field} $F$ in $(1+1)$-dimensions (or two chiral massless fields in the Yang-Mills-Higgs case). The classical partition functions will include a sum of 
\begin{equation*}
e^{-\beta (E-\mu _{B}B)}\ ,
\end{equation*}%
(where $E$ is the energy of the solution and $B$ is the Baryonic charge) over all the possible discrete labels and (for any choice of the discrete labels) over the chiral massless field $F$ satisfying the boundary conditions defined in the previous sections corresponding to the given choice of discrete labels.\footnote{%
The main difference between the Yang-Mills-Higgs and Skyrme cases is that in
the former, two chiral modes contribute to the total energy, while in the
latter, only one.} On the other hand, we can take advantage of the fact that
the massless chiral field $F$ satisfies a linear equation. This allows to
``promote" the classical partition function over $F$ to a ``semi-classical"
partition function by quantizing the massless chiral degree of freedom $F$
in the obvious way.\footnote{%
A more detailed treatment of the partition functions associated with these
families will appear in future publications.}

We will focus mainly on the Skyrme theory since it is more directly relevant as far as the low temperatures phase diagram is concerned (being the Skyrme theory, the low energy limit of QCD at leading order in the 't Hooft expansion). The relations with the instantons-dyons liquid approach 
\cite{PoissonDuality0,PoissonDuality1,Ramamurti:2018evz,PoissonDuality3}
will be shortly analyzed. A complete treatment of the quantum partition functions associated with these families should include (for any member of these families) the other possible fluctuations (such as small perturbations of the other two degrees of freedom of the Skyrme model and not just of $F$). Unfortunately, this task would involve the computation of functional determinants in $(3+1)$-dimensional backgrounds, which depend explicitly on time and spatial coordinates: such a computation can be done neither analytically nor numerically. However, it is worth emphasizing that it is already a quite remarkable fact that one of the modes (namely $F$) can be quantized exactly. Moreover, the comparison with \cite{PoissonDuality0,PoissonDuality1,Ramamurti:2018evz,PoissonDuality3} here below clearly shows that the partition function to be defined in the following sections captures much relevant information.

Schematically, the contribution of the current families of exact solutions
to the partition function $Z$ is%
\begin{equation*}
Z\ \approx \ \sum_{\substack{ over\ all\ the  \\ solutions  \\ of\ the\
family }}\ \exp \left[ -\beta \left( E_{\text{Cl. Sol.}}-\mu _{B}B_{\text{%
Cl. Sol.}}\right) \right] \,,
\end{equation*}%
where the sum is over all the solutions of the given family.\footnote{Here the sum over all the solutions of the family means a sum over $p$, $q$ and $\widetilde{F}$.}
Here $E_{\text{Cl. Sol.}}$ is the total energy of a classical solution, $B_{\text{Cl. Sol.}%
} $ is the Baryonic charge of the configuration, $\beta $ is the inverse of
the temperature $T$, and $\mu _{B}$ is the Baryon chemical potential.

In particular, for the lasagna phase constructed from the Euler
representation and for the spaghetti phase constructed from the exponential
representation in the Skyrme model, the expressions for $E_{\text{Cl. Sol.}}$
in Eqs. \eqref{E2} and \eqref{E3} can be written, respectively, as \begin{equation}
E_{\text{Cl. Sol.}}^{(2)} := E^{(2)} = \widetilde{\Gamma} ^{(2)}+\Psi
^{(2)}\,\int_{0}^{2\pi }\left[ \left( \frac{\partial \widetilde{F}}{\partial
t}\right) ^{2}+\frac{1}{L_{\phi }^{2}}\left( \frac{\partial \widetilde{F}}{%
\partial \phi }\right) ^{2}\right] d\phi \,,  \label{etot2}
\end{equation}
where 
\begin{equation*}
\widetilde{\Gamma} ^{(2)} = \Gamma^{(2)}+ \frac{64 \pi q^2}{L_{\phi}^{2}}
\Psi^{(2)} \ ,
\end{equation*}
and 
\begin{equation}
E_{\text{Cl. Sol.}}^{(3)} := E^{(3)} = \widetilde{\Gamma} ^{(3)}+\Psi
^{(3)}\,\int_{0}^{2\pi }\left[ \left( \frac{\partial \widetilde{F}}{\partial
t}\right) ^{2}+\frac{1}{L_{\phi }^{2}}\left( \frac{\partial \widetilde{F}}{%
\partial \phi }\right) ^{2}\right] d\phi \,,  \label{etot3}
\end{equation}
where 
\begin{equation*}
\widetilde{\Gamma} ^{(3)} = \Gamma^{(3)}+ \frac{4 \pi p^2}{L_{\phi}^{2}}
\Psi^{(3)} \ .
\end{equation*}
Here $\{\Gamma ^{(2)},\Psi ^{(2)}\}$ and $\{\Gamma ^{(3)},\Psi ^{(3)}\}$
have been defined, respectively, in Eqs. \eqref{E2} and \eqref{E3}, and $\widetilde{F}$ is
the part of $F$ coming from the sum over the integers $n$ in Eqs. (\ref%
{general1}) and (\ref{general2}). It is important to remember that the
linear terms in Eqs. \eqref{etot2} and \eqref{etot3} that comes from the
modes expansion of the function $F$ must be non-zero in order to have a
non-vanishing topological charge. Also, it is worth emphasizing that the novel solutions presented in the
manuscript with the arbitrary dependence of $F$ on $u$ (which is the
light-like coordinate orthogonal to the space-like coordinates which enter
explicitly in the ansatz) can be considered as saddle points: from the
intuitive viewpoint the periodic part of $F(u)$ represents the lowest energy
normal modes of the Hadronic tubes and layers in very much the same way as a
vibrating string encodes the lowest energy normal modes of a static strings.

\subsection{Partition function for fixed value of the Baryonic charge}

In the following, we will focus on the nuclear lasagna phase, which is slightly simpler to analyze than the spaghetti phase, using as starting point Eq. \eqref{etot2}. The reason is that the total and free energies associated with Hadronic tubes depend on inverse elliptic functions, while the ones of Hadronic layers depend on functions which are polynomial in the physically relevant variables (so that these are easier to handle). On the other hand, the qualitative low-temperature behaviour for large Baryonic charges is similar in both cases.

Let us consider a fixed value of the Baryonic charge $B$ in Eq.~%
\eqref{topchargeNLSM} and let us turn off, momentarily, the Baryon chemical
potential $\mu _{B}$. 

In order to avoid very long algebraic expressions we will consider $q=p$ (since this choice keeps the essential features of the problem). 
As for fixed values of the discrete label $p$ these configurations are characterized by a massless chiral field $F$
in two dimensions, the contribution $Z_{p}$ of the present family to the
Skyrme partition function is 
\begin{equation}
Z_{p}(\beta)=\int DF\mathcal{Z}_{F}=\int DF\exp \left\{ -\beta E_{\text{%
Cl. Sol.}}^{(2)}\right\} \,,~~~~~\mathcal{Z}_{F}=\exp \left\{ -\beta E_{%
\text{Cl. Sol.}}^{(2)}\right\} \,,  \label{partition4}
\end{equation}%
where $E_{\text{Cl. Sol.}}^{(2)}$ have been defined in Eq.~\eqref{etot2}
(the case of Hadronic tubes, defined by the energy in Eq.~\eqref{etot3}, has
a similar qualitative behaviour). The path integral over the massless chiral
field can be done in the usual way taking into account the obvious
quantization (see, for instance, \cite{GSW}) of the mode expansion for $F$
(here, we will consider the total Baryonic charge to be even to avoid
complications with Grassmann variables associated to $\phi _{0}^{-}$ and $%
v_{-}$):%
\begin{equation}
F_{-}=\phi _{0}^{-}+v_{-}\left( \frac{t}{L_{\phi }}-\phi \right)
+\sum_{n\neq 0}\left( a_{n}^{-}\sin \left[ n\left( \frac{t}{L_{\phi }}-\phi
\right) \right] +b_{n}^{-}\cos \left[ n\left( \frac{t}{L_{\phi }}-\phi
\right) \right] \right) \ .  \label{partition4.9}
\end{equation}%
The corresponding semi-classical partition function for the Hadronic layers
(with fixed discrete labels) reads: 
\begin{equation}
Z_{p}\left( \beta \right) =\sum_{n=1}^{+\infty }\delta \left( n\right)
\exp \left[ -\beta \left( \widetilde{\Gamma}^{(2)}+\Psi ^{(2)}n\right) %
\right] \,, \label{partition4.99}
\end{equation}%
where the integer $n$ comes from the quantization of the Hamiltonian, $%
\displaystyle \int_{0}^{2\pi }\left[ \left( \frac{\partial \widetilde{F}}{\partial t}%
\right) ^{2}+\frac{1}{L_{\phi }^{2}}\left( \frac{\partial \widetilde{F}}{%
\partial \phi }\right) ^{2}\right] d\phi $, of the massless chiral mode and $%
\delta (n)$ is the corresponding degeneracy (related to the number
partition). Taking into account Eq.~\eqref{energy2}, one can rewrite $%
\widetilde{\Gamma}^{(2)}$ and $\Psi ^{(2)}$ as follows: 
\begin{gather}
\widetilde{\Gamma} ^{(2)} = \Sigma _{1}+\Sigma _{2}p^{2} +\Sigma _{3}p^{4}\
,\quad \Psi ^{(2)}=\Sigma _{4}+\Sigma _{5}p^{2}\ , \quad \Sigma _{1}=\frac{%
K\pi ^{3}L_{\theta }L_{\phi }}{8L_{r}} \,,  \notag \\
\Sigma _{2} = \frac{K\pi ^{3}}{32L_{r}L_{\theta } L_{\phi}}(\lambda+16L_{r}^{2})(L_{
\phi}^{2}+32 L_{\theta}^{2}) \ , \quad \Sigma_{3}= \frac{8K\pi^2 L_{r} \lambda}{L_{\theta}
L_{\phi}} \ , \quad \Sigma _{4}=\frac{K\pi ^{2}L_{\theta }L_{\phi }}{64L_{r}}%
\left( \lambda +16L_{r}^2\right) \ , \quad \Sigma _{5}=\frac{K \lambda \pi
L_{r} L_{\phi }}{8 L_{\theta }}\,.  \label{usefulabel}
\end{gather}%
The above is useful to separate the terms which depend on the discrete
labels (which are proportional to $\Sigma _{2}$, $\Sigma_{3}$ and $\Sigma
_{5}$) from the terms which do not depend on any discrete label of the
family (which are proportional to $\Sigma _{1}$ and $\Sigma _{4}$).
Note that the partition function in Eq. (\ref{partition4.99}), when $p\neq q$, will be given by
\begin{equation*}
Z_{p,q}\left( \beta \right) =\exp \left[ -\beta \left( \Sigma _{1}+\Sigma
_{2}\frac{(L_\phi^2 p^{2}+32 L_\theta^2 q^2)}{(L_\phi^2+32 L_\theta^2)} + \Sigma_{3} p^2 q^2 \right) \right] \sum_{n=1}^{+\infty }\delta
\left( n\right) \exp \left[ -\beta \left( \left( \Sigma _{4}+\Sigma
_{5}p^{2}\right) n\right) \right] \ . 
\end{equation*}
Now, in our case (with $p=q$), Eq. (\ref{partition4.99}) can be written as
\begin{equation}
Z_{p}\left( \beta \right) =\exp \left[ -\beta \left( \Sigma _{1}+\Sigma
_{2}p^{2} + \Sigma_{3} p^4 \right) \right] \sum_{n=1}^{+\infty }\delta
\left( n\right) \exp \left[ -\beta \left( \left( \Sigma _{4}+\Sigma
_{5}p^{2}\right) n\right) \right] \ .  \label{partition5.01}
\end{equation}
These results are similar to the usual two-dimensional chiral CFT with the difference that, in the sum over $n$, the inverse temperature $\beta $ has been rescaled by $\left( \Sigma _{4}+\Sigma _{5}p^{2}\right) $. If $p$ is fixed, then the result is the usual chiral massless bosons partition function with rescaled temperature $\beta _{r}:=\left( \Sigma _{4}+\Sigma
_{5}p^{2}\right) \beta $. However, beware that we also have to sum over the label $p$. Such a partition function can also be written as 
\begin{equation}
\zeta \left( z\right) \sim \exp \left( -\beta \left( \Sigma _{1}+\Sigma
_{2}p^{2}+\Sigma_{3} p^4 \right) \right) \sum_{n=1}^{+\infty }\delta \left(
n\right) \exp (-\beta _{r}n)\,,  \label{partition6.1}
\end{equation}
where $\delta (n)$ is the degeneracy of the energy level $n$ which can be
easily obtained (for large $n$) using the Hardy-Ramanujan-Cardy formula. The
fundamental formula for the asymptotic growth of the partitions $\delta (n)$
was found long time ago by Hardy and Ramanujan in Ref. \cite%
{ramanujanformula}: for $n\gg 1$ we get 
\begin{equation}
\delta \left( n\right) \sim \frac{1}{4\sqrt{3}}\frac{\exp \left( \pi \sqrt{%
\frac{2n}{3}}\right) }{n}\,.  \label{ramanujaneq}
\end{equation}
Such discrete label $n$ represents \textit{exact excitation energies}\footnote{%
We use the expression \textquotedblleft exact excitation energies", since these chiral modes are not only \textquotedblleft small excitations" on top of Hadronic tubes or layers but, in fact, \textit{these configurations are exact solutions of the full Skyrme field equations}. On the other hand, in the usual cases, one can only study small fluctuations around topological solitons as \textit{solutions of the linearized field equations around that
solitons}.} (related to the \textquotedblleft quanta" of modulations either
of the Hadronic layers or of the Hadronic tubes) over \textquotedblleft
bare" lasagna or spaghetti configurations. Looking at Eq. (\ref{partition4.9}%
), the \textquotedblleft bare" Euler or Exponential configurations (which
have been discussed previously in the literature, see \cite%
{gaugsk,LastUS1,LastUS2}, \cite{Sergio1,Sergio2} and references therein)
possess $a_{n}=0=b_{n}$ while $v_{-}\neq 0$ (and fixed by the boundary
condition to have integer Baryonic charge). Since Skyrme theory is the low
energy limit of QCD, it is natural to introduce a cut-off $\Delta $ on the
sum over $n$: such a $\Delta $ can be interpreted as the scale beyond which
the Skyrme model is not a good description anymore. Therefore, instead of
Eq. (\ref{partition6.1}), we will consider the following expression: 
\begin{equation}
\zeta _{\Delta }\left( z\right) \sim \exp \left( -\beta \left( \Sigma
_{1}+\Sigma _{2}p^{2} + \Sigma_{3} p^4 \right) \right) \sum_{n=1}^{\Delta
}\delta \left( n\right) \exp (-\beta _{r}n)\,.  \label{partition6.2}
\end{equation}
The $\Delta $ depends, in general, both on the temperature and the chemical
potential: $\Delta =\Delta (T,\mu _{B})$. Although we have been unable to find in the literature a widely accepted
expression for the cut-off $\Delta =\Delta (T,\mu _{B})$ as function of $T$
and $\mu _{B}$, in the following subsection, we will show that reasonable
choices of $\Delta $ provide with analytic results in qualitative agreement
both with the available lattice data \cite{Caselle:2015tza} as well as with
different analytical approach \cite{Ramamurti:2018evz}.

\subsection{Partition function at finite Baryon chemical potential}


In order to get the full contribution of the present family to the
semi-classical Skyrme partition function with non-vanishing Baryon chemical
potential $\mu _{B}$ one also has to sum over $p$, since $p$
determines the Baryon charge $B=p^2$ (remember that we have considered for simplicity $q=p$). In this way, we get
\begin{equation*}
Z^{\ast }=\sum_{p=-\infty }^{+\infty }\exp \left( -\beta \left( \Sigma
_{1}+\left( \Sigma _{2}-\mu _{B}\right) p^{2} +\Sigma_{3}p^4 \right) \right)
\sum_{n=1}^{\Delta }\delta \left( n\right) \exp (-\beta_r n )
\label{partition7}
\end{equation*}%
\begin{equation}
\Leftrightarrow Z^{\ast }=\sum_{n=1}^{n_{\max }}\sum_{p=-\infty }^{+\infty
}\delta \left( n\right) \exp \left( -\beta \left( \Sigma _{1}+\left( \Sigma
_{2}-\mu _{B}\right) p^{2} +\Sigma_{3} p^4\right) \right) \exp (-\beta
\left( \Sigma _{4}+\Sigma _{5}p^{2}\right) n )\,,~~~~~~n_{\max }=\Delta
\left( T,\mu _{B}\right) \,.  \label{jacobi}
\end{equation}%
The above double sum is clearly convergent since it is possible to exchange
the order of the sums. 
As we will see below, the cut-off $n_{\max }=\Delta \left( T,\mu _{B}\right) 
$ can be fixed in such a way to achieve a qualitative agreement with the
description of LQCD for the phase diagram. Note that, unlike what happens in
LQCD, in the present approach the inclusion of the Baryon chemical potential
is not harmful.

It is worth emphasizing the intriguing similarities of the present partition
function with the semi-classical partition functions computed using the
\textquotedblleft Poisson duality" and the instanton-dyon liquid approach in
SUSY Yang-Mills theory (see \cite%
{PoissonDuality0,PoissonDuality1,Ramamurti:2018evz,PoissonDuality3} and
references therein). Let us consider first the case in which $\frac{L_{r}}{%
L_{\phi }}$ is very small $\left(\frac{L_{r}}{L_{\phi }}\ll 1\right)$ so that $\Sigma
_{3}$ can be neglected (see Eq.~\eqref{usefulabel}). In this case (which
corresponds to a box which is much longer in the $\phi $-direction than in
the $r$-direction), if one analyzes Eq.~(11) at page $5$ of Ref. \cite%
{Ramamurti:2018evz}, one can see that the label $k$ (and the corresponding
sum) of $Z_{inst}$ in that reference is analogous to the sum over $n$ in
Eq. \eqref{jacobi} in the present approach,
as $k$ appears linearly in the exponent of $Z_{inst}$ as $n$ in Eq. \eqref{jacobi}.
On the other hand, the label $n$
of $Z_{inst}$ (and the corresponding sum) in Eq. (11) of Ref. \cite%
{Ramamurti:2018evz} is analogous to our topological sums over $p$ since the
label $n$ appears quadratically in the exponent of $Z_{inst}$ of Ref. \cite%
{Ramamurti:2018evz}, as $p$ in our case. The only two relevant differences
between the present expressions and $Z_{inst}$ are the following. \textit{%
First}, in the sum in Eq. (11) of Ref. \cite{Ramamurti:2018evz} there is the
factor $\left( \beta /g^{2}\right) \left( k^{3}/\beta M\right) ^{3}$ where $M
$ is defined below Eq. (11) of Ref. \cite{Ramamurti:2018evz} while in our
case we have the degeneracy factor $\delta \left( n\right) $. The factor
arises from the one-loop effects around the instantons and can be computed
explicitly thanks to the powerful results made available by SUSY, which are
basic building blocks in the approach introduced in \cite%
{PoissonDuality0,PoissonDuality1}. However, in the low energy/temperature
limit of QCD, there is no SUSY, so the computations of one-loop effects are
far more complicated. It is worth reminding that each term in the expansion
in Eq.~\eqref{partition4.9} corresponds to an exact solution of the Skyrme
field equations (with energy and Baryon densities depending on all the three
spatial coordinates in a non-trivial way), so that, for any fixed $n$ in Eq.
(\ref{jacobi}), one should compute the corresponding 1-loop determinant
around this non-trivial non-supersymmetric background. This fact, together
with the lack of SUSY, makes the computation of this 1-loop determinant
unfeasible in our case. \textit{Second}, in the present approach, we have
introduced a cut-off on the sum over $n$ as the Skyrme model is not valid
anymore at very high temperature/energies while SUSY Yang-Mills theory is
well behaved in the UV. Despite these differences, we find the similarities
between the two approaches quite striking. On the other hand, if $\frac{L_{r}%
}{L_{\phi }}$ is not very small, then the thermodynamical behavior of
present families of topologically non-trivial configurations will deviate
from the predictions of the instanton-dyon liquid approach (when the
adimensional parameter $\frac{L_{r}}{L_{\phi }}$\ plays a key role). The
very rich but quite complicated phase diagram associated to these families
will be analyzed in a future publication.

The idea of the present section is to provide sound pieces of evidence that the families of topologically non-trivial configurations constructed in the previous sections have a reasonable thermodynamical behaviour. In order to get an idea of the thermodynamical behaviour of these modulated topological
solitons, we can approximate the sums in Eq.~\eqref{jacobi} by integrals
(in the limit in which$\frac{L_{r}}{L_{\phi }}\ll 1$ so that $\Sigma _{3}$
can be neglected: see Eq. (\ref{usefulabel})), arriving at the following
formula 
\begin{align}\label{pequaltoq}
\nonumber Z_{GP}(\widetilde{\mu }_{B},T)=& ~\exp \left( -\beta \Sigma _{1}\right)
\int_{1}^{\Delta (T,\mu _{B})+1}%
\hspace{-0.3cm}dn\, \delta (n) \exp \left(-\beta \Sigma _{4}n\right)\, \int_{-\infty }^{+\infty }\hspace{-0.2cm}dp\,\exp \left( -p^{2}\beta (\Sigma _{5}n-%
\widetilde{\mu }_{B})-\beta \Sigma_{3} p^4\right)   \\
=&~\frac{\exp \left( -\beta \Sigma _{1}\right)}{2 \sqrt{\Sigma_{3}}}\, \int_{1}^{\Delta (T,\mu _{B})+1} \hspace{-0.9cm}dn \, \delta (n) \,\exp \left( \beta \left[\frac{ (n \Sigma_{5}-\widetilde{\mu }_{B})^2}{8\Sigma_{3}} - \Sigma _{4}n \right]\right) \,  \sqrt{n \Sigma_{5}- \widetilde{\mu }_{B}}\, K_{1/4} \left( \frac{\beta (n \Sigma_{5}-\widetilde{\mu }_{B})^2}{8\Sigma_{3}} \right) \ , 
\end{align}
where $\widetilde{\mu }_{B}:=\mu _{B}-\Sigma _{2}$, $K_{n}(z)$ denotes the modified Bessel function of the second kind, and $\delta (n)dn$ gives the number of states
with energies between $n$ and $n+dn$. Note that the condition $\Sigma_2+\Sigma_5>\mu_B$ must be fulfilled. Also, we have introduced a $+1$ in the upper integration limit of $n$ for numerical analysis reasons.\\

The integral in Eq. \eqref{pequaltoq} cannot be computed exactly. Since we have to evaluate it numerically, we can consider the following generalized form of $\delta (n)$. For this section, let us consider a modified expression of
Eq. \eqref{ramanujaneq}, as follows 
\begin{equation}
\delta (n)\sim \frac{1}{4\sqrt{3}}\frac{\exp \left( \sqrt{\frac{2n}{3}}\pi
\right) }{(n^{a}+b^{2})}\ ,~~~~~~~\text{as}~~~n\rightarrow \infty
\,,~~~~~a,b\in \mathbb{R}_{>0}\,.  \label{rama}
\end{equation}%
The original formula in Eq. \eqref{ramanujaneq} recovers by setting $a=1$
and $b=0$. Substituting Eq.~\eqref{rama} into Eq. \eqref{pequaltoq}, we get
\begin{equation}\label{mainZ}
Z_{GP}(\widetilde{\mu }_{B},T)\approx \frac{\exp \left( -\beta \Sigma _{1} +\sqrt{\frac{2n}{3}}\pi\right)}{8 \sqrt{3 }}\, \int_{1}^{\Delta (T,\mu _{B})+1} \hspace{-0.2cm}dn \, f(n,\mu _{B})\,,
\end{equation}
where
\begin{equation}
f(n,\mu _{B})=\frac{\sqrt{ n \Sigma_{5}- \widetilde{\mu }_{B}}}{\sqrt{ \Sigma_{3}}(n^{a}+b^{2})}\, \exp \left( \beta \left[\frac{ (n \Sigma_{5}-\widetilde{\mu }_{B})^2}{8\Sigma_{3}} - \Sigma _{4}n \right]\right) \, K_{1/4} \left( \frac{\beta (\widetilde{\mu }_{B}-n \Sigma_{5})^2}{8\Sigma_{3}} \right)\,.
\end{equation}
By considering the expansion $L_{r} /L_{\phi }\ll 1$, the last function reduces to
\begin{equation}
f(n,\mu _{B}) \approx  \frac{2\sqrt{\pi} e^{-n\beta \Sigma_4 }}{ (n^a+b^2) \sqrt{ \beta(\widetilde{\mu }_{B}-n \Sigma_{5})}}\,, ~~~\text{as} ~~~~ \Sigma_3 \ll 1\,.
\end{equation}
The partition function in Eq.~\eqref{mainZ} with the expansion $\Sigma_3 \ll 1$ allows one to extract
different thermodynamical properties of the present families of
\textquotedblleft dressed topological solitons": we will compare the results
obtained from the above partition function with the available numerical
results from LQCD. Before doing that, we should note that the explicit
dependency on the temperature in the limit of integration through the
cut-off $\Delta (T,\mu _{B})$ can also be considered as a modification in
the Hamiltonian with additional explicitly $T$-dependent terms. As it was
shown by Gorenstein and Yang \cite{Gorenstein:1995vm}, this kind of
modification produces specific changes in some thermodynamics functions. At
finite chemical potential, a generalization of the solution of Gorenstein
and Yang modifies the entropy $S$ and the internal energy $U$ as \cite%
{Gardim:2009mt} 
\begin{eqnarray}
S^{\prime }(V,T,\Delta (T,\mu _{\scriptscriptstyle B})) &\equiv
&S(V,T,\Delta (T,\mu _{\scriptscriptstyle B}))-\frac{\partial \Delta }{%
\partial T}\left( \frac{\partial A}{\partial \Delta }\right) _{V,T}\,,
\label{entropyGardim} \\
U^{\prime }(V,T,\Delta (T,\mu _{\scriptscriptstyle B})) &\equiv
&U(V,T,\Delta (T,\mu _{\scriptscriptstyle B}))-T\,\frac{\partial \Delta }{%
\partial T}\left( \frac{\partial A}{\partial \Delta }\right) _{V,T}\,,
\end{eqnarray}%
where $A(V,T)$ is identified with the free energy obtained from the standard
formula 
\begin{equation}
A(\mu _{B},T,V)=-T\log Z_{GP}(\mu _{B},T,V)\,.
\end{equation}%
All the other thermodynamics functions are unchanged and can be found from $%
A(\mu _{B},T,V)$ using the standard thermodynamics relations. In order to
compare our simulations with LQCD, we are particularly interested in
computing the pressure 
\begin{equation}
P=-\left( \frac{\partial A}{\partial V}\right) _{T}\,,
\end{equation}%
with $V=8\pi^3 L_{r} L_{\theta} L_{\phi}$ being the finite volume. In the limit of a $T$-independent
Hamiltonian the last term in Eq. \eqref{entropyGardim} has to be zero, so
that the standard expressions of the statistical mechanics are recovered.

As it is well known, perturbative QCD calculations should describe, at
extremely high temperatures and chemical potential, the quarks and gluons
degrees of freedom: the quark-gluon plasma. To make our results comparable
with perturbative QCD computations, at those energies, we will also add to
the pressure and entropy the perturbative terms, to order $g^{2}$, computed
in the perturbative QCD approach \cite{Blaizot:1999ip,Blaizot:2003tw}, given
by 
\begin{eqnarray}
S^{\prime }(V,T,\Delta (T,\mu _{B})) &=&S(V,T,\Delta (T,\mu _{B}))-\frac{%
\partial \Delta }{\partial T}\left( \frac{\partial A}{\partial \Delta }%
\right) _{V,T}+\left( \frac{\pi ^{2}(7N_{c}N_{f}+4N_{g})}{45}-\frac{%
N_{g}(4N_{c}+5N_{f})}{144}g^{2}\right) T^{3}\,,  \label{pressureeq} \\
P &=&-\left( \frac{\partial A}{\partial V}\right) _{T}+\left( \frac{\pi
^{2}(7N_{c}N_{g}+4N_{g})}{180}-\frac{N_{g}(4N_{c}+5N_{f})}{576}g^{2}\right)
T^{4}\,,
\end{eqnarray}%
where $g$ is the coupling constant of QCD, $N_{g}=(N_{c}^{2}-1)$, $N_{c}$ is
the number of colors, and $N_{f}$ is the number of flavors.

A possible choice of the cut-off $\Delta (T,\mu _{B})$ at a fixed chemical
potential that allows comparing closely our results to LQCD data is 
\begin{equation}
\Delta (T,\mu _{B})=0.050\times \log \left(
0.712-2.148T+0.294T^{2}+79.971T^{-1}+14.254T^{-2}+25.413e^{-1.605T}\right)
\,,
\end{equation}%
where we have fixed $\mu _{B}=0.1$. In the present case the temperature has energy units (our energy unit is $100 \text{MeV}$ as it is common in QCD), so that the first term inside the logarithmic is dimensionless, the second coefficient has inverse energy units [$2.148 \times (100 \text{MeV})^{-1}$], and so on.
On the other
hand, it would be nice to determine the precise functional form of $\Delta
(T,\mu _{B})$\ from first principles: we hope to come back on this
interesting issue in a future publication. In the meantime, our aim is only
to show that the partition function associated to the configurations
described in the previous sections can be relevant as simple choices of $%
\Delta (T,\mu _{B})$ and give rise to good qualitative agreement with LQCD.

\begin{figure}[h!]
 \centering
 \includegraphics[width=.4\linewidth]{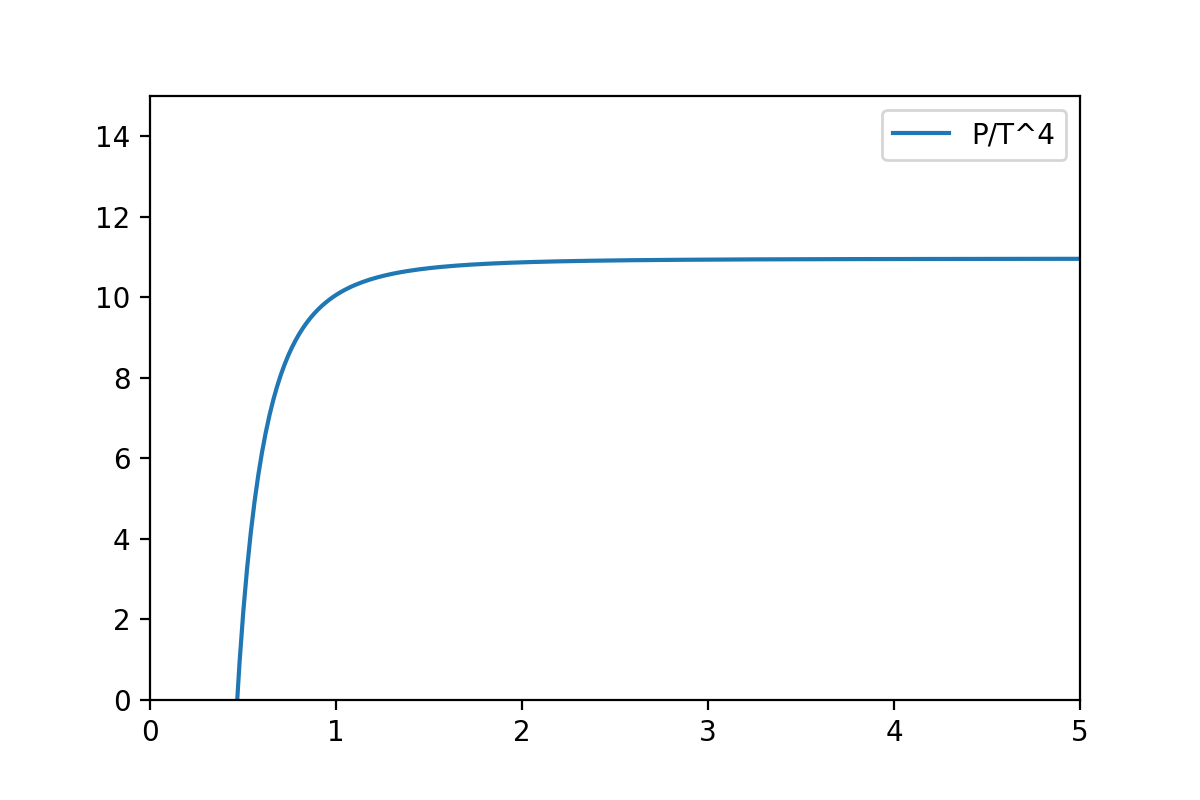}
  \includegraphics[width=.4\linewidth]{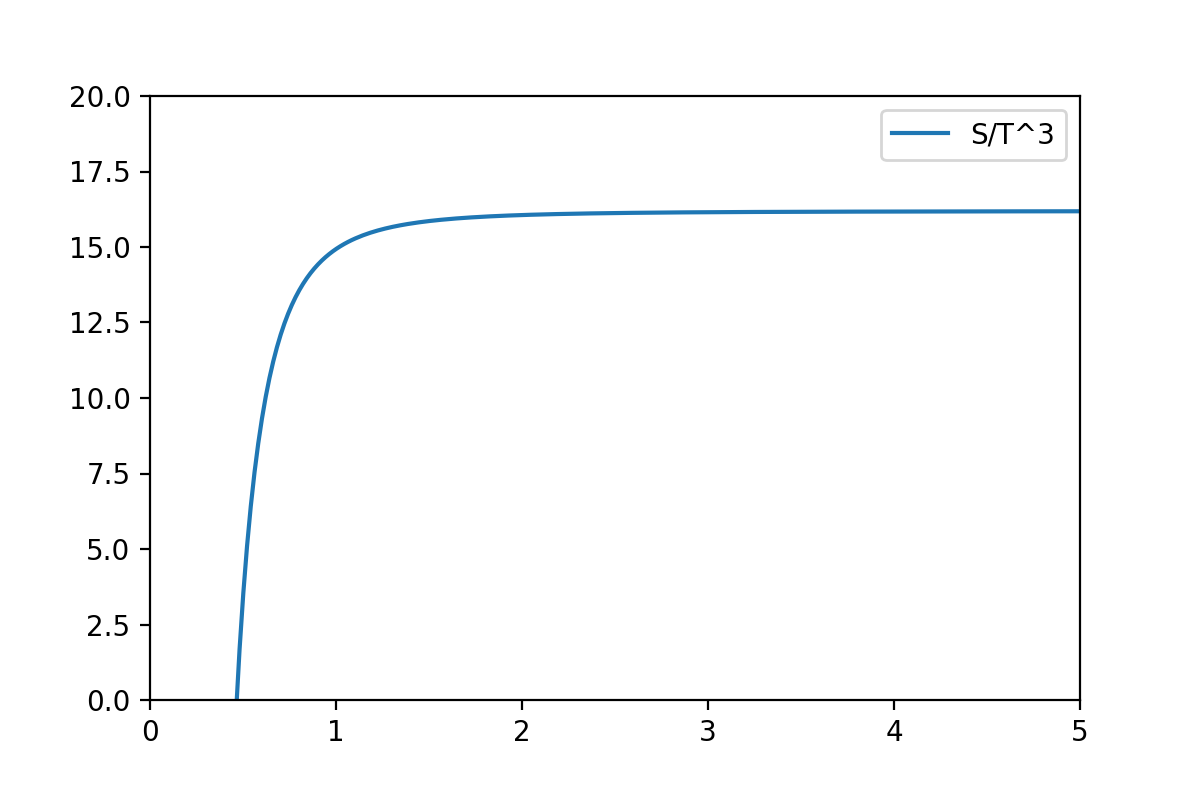}
\caption{Plots of the pressure normalized by $T^{4}$ and the entropy
normalized by $T^{3}$ as functions of the temperature $T$. We run our
simulations with the values $\protect\mu _{\scriptscriptstyle B}=0.1$, $V=1$%
, $N_{g}=N_{c}^{2}-1$, $N_{c}=3,N_{f}=2,g=0.1,a=1,b=0$.}
\label{fig:fig}
\end{figure}

Now, one of the primary thermodynamic observables that we compute is the
pressure $P$ according to the formula in Eq. \eqref{pressureeq}. With the
expression of $\Delta (T,\mu _{B})$ here above, the starting value for $%
P/T^{4}$ at $T_{in}\equiv 0.480$ is given by $P_{in}/T_{in}^{4}=0.942$. The results at different values of $T$ are
shown in Fig.~\ref{fig:fig}. The comparison of this plot can be made with
those from Refs. \cite{Giusti:2016iqr,Borsanyi:2012ve,Caselle:2018kap}.
Another crucial thermodynamic observable of our interest is the entropy $S$
related to the pressure by basic thermodynamics relations. The starting
point of the entropy at $T_{in}=0.480$ is $S_{in}/T_{in}^{3}=1.679$. The entropy per unit of $T^{3}$ is shown in
Fig.~\ref{fig:fig}. 

Clearly, these plots exhibit good qualitative agreement with the results of
LQCD of those references. We will come back on a more detailed analysis of
the low temperature behavior of the present topologically non-trivial
configurations in a future publication.


\section{Conclusions and perspectives}


In the present work, we constructed exact and topologically non-trivial
solutions of the Skyrme and Yang-Mills-Higgs theory at finite Baryon density
in $(3+1)$ dimensions. These analytic configurations are characterized both by two discrete labels (determining the Baryonic charge) and by a massless
chiral field $F$ in $(1+1)$ dimensions (in the Yang-Mills-Higgs case, there
are two chiral massless modes). Physically, the chiral massless
modes characterize exact excitations on top of Hadronic layers and tubes.
Thus, non-trivial modes of $F$ represent either Hadronic tubes which are not
homogeneous along the axis of the tubes or Hadronic layers, which are not
homogeneous in the directions tangent to the layers themselves. In other
words, the chiral massless modes hosted in the topologically non-trivial
configurations constructed in the previous sections represent
\textquotedblleft exact excitations" since these chiral modes are not only
\textquotedblleft small excitations" on top of Hadronic tubes or layers but,
\textit{these configurations are exact solutions of the full Skyrme
field equations with non-trivial topological density (the same is true in
the Yang-Mills-Higgs case)}. This situation should be compared with the
usual circumstances when one can only study small fluctuations around topological
solitons as \textit{solutions of the linearized field equations around that
solitons}. Hence, these are the first exact analytic examples describing
ordered arrays of $(3+1)$-dimensional topological solitons with non-trivial
inhomogeneities. In the case of the Skyrme model, the plots of the energy
and Baryon densities of the two types of solutions show that these
configurations are appropriate to describe inhomogeneous nuclear pasta
states, where chiral modes modulate the tubes and layers.

From the technical viewpoint, the fact that the present approach can reduce
the complete set of field equations of the Skyrme model in $(3+1)$-dimensions to the equation of a massless chiral field in $(1+1)$-dimensions
(keeping alive the Baryonic charge) open the remarkable possibility to use
tools from CFT in $(1+1)$-dimensions to analyze the low-temperature behaviour of QCD. We have discussed the semi-classical grand canonical
partition function associated with one of the present families. We have
calculated (by approximating the partition function of the Hadronic layer
with a suitable one-dimensional integral) the pressure and the entropy,
obtaining an excellent qualitative agreement with results from LQCD. Our
results also allow discussing out-of-equilibrium features in the low energy
limit of QCD in $(3+1)$-dimensions using the well-established tools of
two-dimensional CFT (see \cite{outofequilibrium}\ and references therein).
We will analyze these issues in a future publication.

\subsection*{Acknowledgments}


F. C. has been funded by Fondecyt Grant No. 1200022. M. L. is funded by
FONDECYT post-doctoral Grant No. 3190873. A.V. is funded by FONDECYT
post-doctoral Grant No. 3200884. The Centro de Estudios Cient\'{\i}ficos
(CECs) is funded by the Chilean Government through the Centers of Excellence
Base Financing Program of ANID.

\end{document}